Technical Report

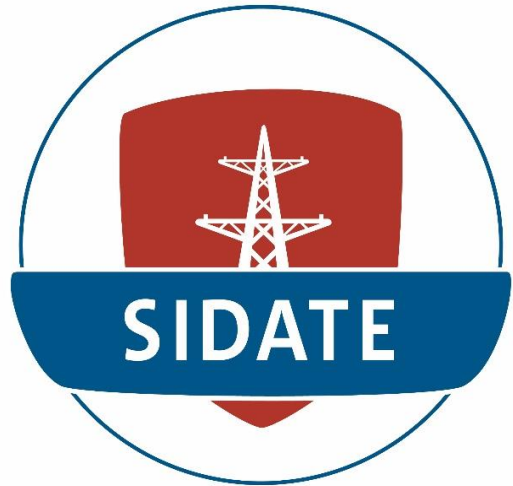

# IT Security Status of German Energy Providers

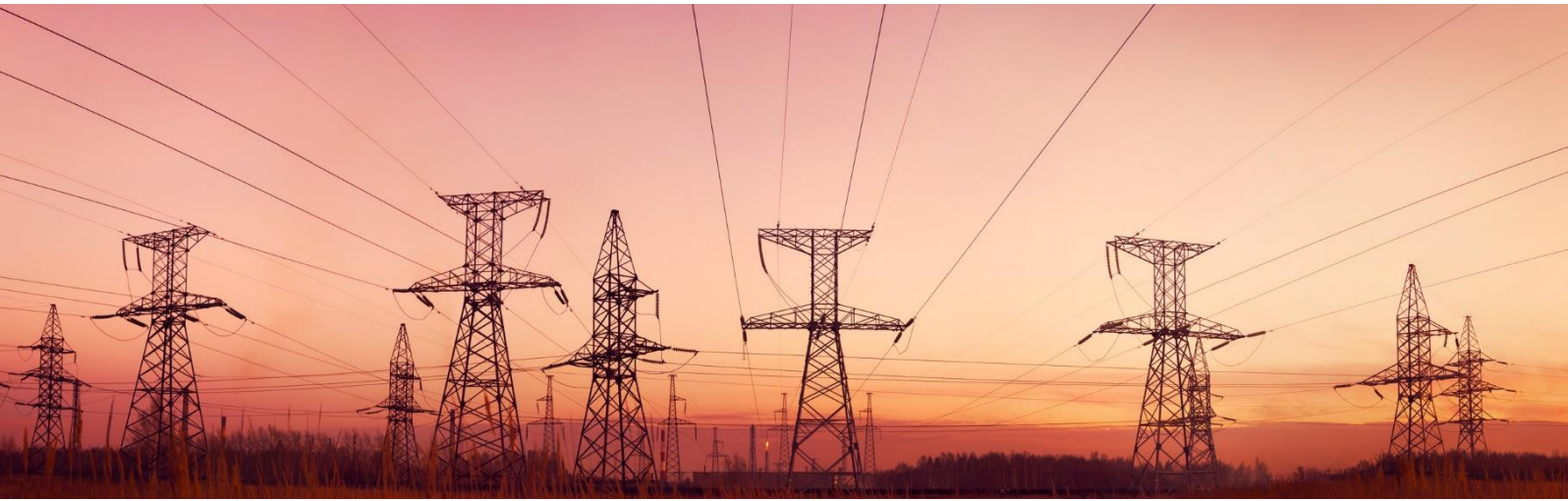

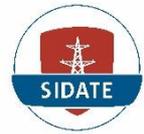

## The following people contributed to this report:

Julian Dax, University of Siegen
Ana Ivan, Goethe University Frankfurt
Benedikt Ley, University of Siegen
Sebastian Pape, Goethe University Frankfurt
Volkmar Pipek, University of Siegen
Kai Rannenberg, Goethe University Frankfurt
Christopher Schmitz, Goethe University Frankfurt
André Sekulla, University of Siegen

## Secure information networks of small- and medium-sized energy providers (SIDATE)

The focus of the SIDATE research project is the technical support of small- and medium-size energy providers for the self-assessment and improvement of their IT security. Different concepts and tools are developed and evaluated by the University of Siegen, Goethe University Frankfurt, regio iT Gesellschaft für Informationstechnologie mbh, and Arbeitsgemeinschaft für sparsame Energie- und Wasserverwendung (ASEW).
More information is available on the project's website http://sidate.org/ .

## Funding

This research project was funded by the Federal Ministry of Education and Research (BMBF) as part of the funding focus "IT Security for Critical Infrastructure".

## Image credit

Image on title: ©TebNad / Fotolia

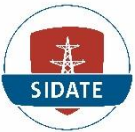

## Table of Contents

## Introduction

As part of the research project "Secure information networks of small- and medium-sized energy providers" (SIDATE), a survey about the IT security status of German energy providers was conducted. The project itself is focused on the IT security of small- and medium-sized energy providers.

In August 2016, 881 companies listed by the Federal Network Agency were approached. Between, September 1st 2016 and October 15th 2016, 61 (6.9%) of the companies replied. The questionnaire focuses on the implementation of the regulatory requirements and on the implementation of an information security management system (ISMS). Additionally, questions about the energy control system, the network structure, processes, organisational structures, and the IT department were asked. Questions were asked in German, so all questions and answers are translated for this report.

Subsequently, the result of the survey is presented. Some questions were only answered by very few participants, and therefore, the related results are not presented.

The survey is organised as follows:
    A) General Company Information
    B) Organisational Aspects
    C) Information Security Management System (ISMS)
    D) Office IT
    E) Energy Control System: Network Structure
    F) Energy Control System: Processes and Organisation

There are two different types of bar charts. The first only contains blue bars. These charts pertain to all energy providers that replied to the specific question. In contrast, the second type of bar charts presents a categorical differentiation between energy providers. The differentiation lies in the size of the company, which is represented by the number of corresponding meter points.

In some cases, spider-web diagrams were used. A categorisation of the responding companies was once again not included in these cases.

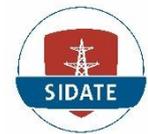

# Part A: General Company Information

In the first part of the survey, general questions were asked in order to present an overview of the participating energy providers. Based on the replies, the energy providers were classified in four overarching categories, in order to achieve a better analysis of the following survey items.

The categorization of the participants uses the number of meter points of the providers as a criterion. The distribution by size is depicted in Figure 1. For the remaining analyses, the providers are separated into the following categories: small (between 0 and 15,000 meter points), medium (between 15,001 and 30,000 meter points), large (between 30,001 and 100,000 meter points), and very large (more than 100,001 meter points).

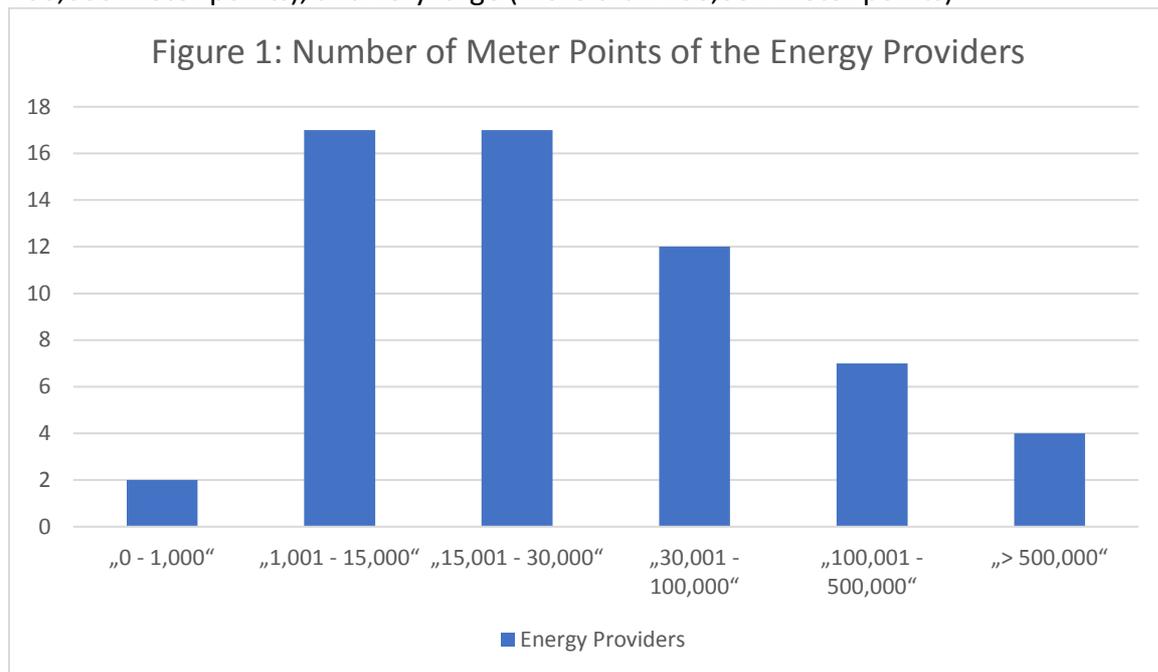

Figure 1: How many meter points are in your network?

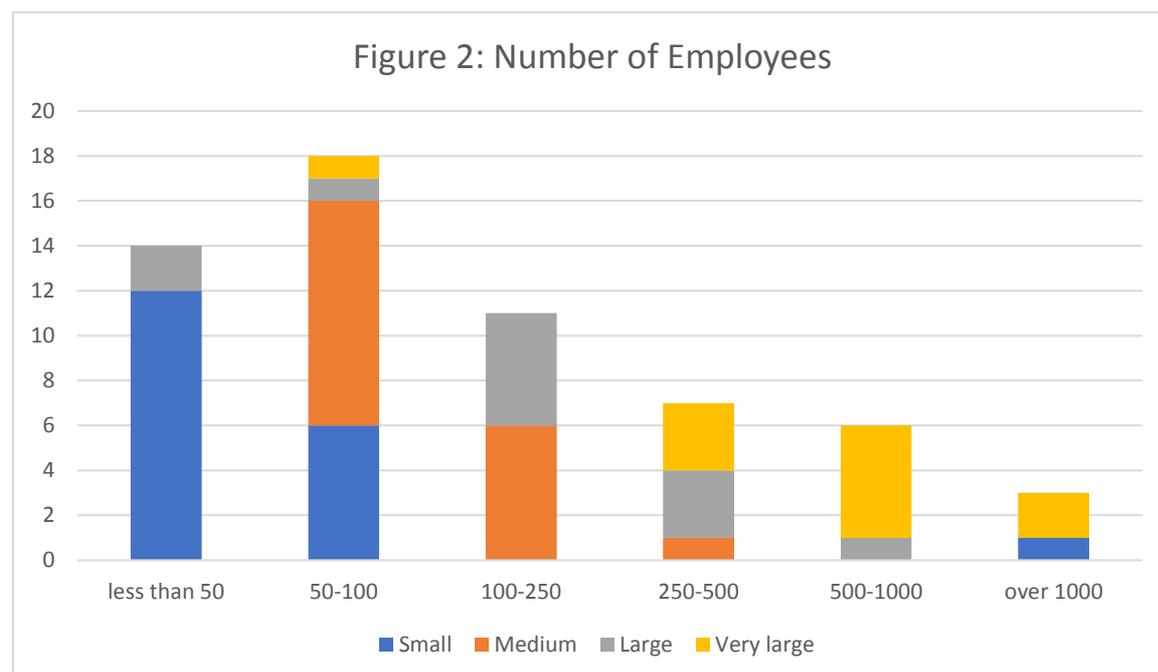

Figure 2: How many employees are in your company?

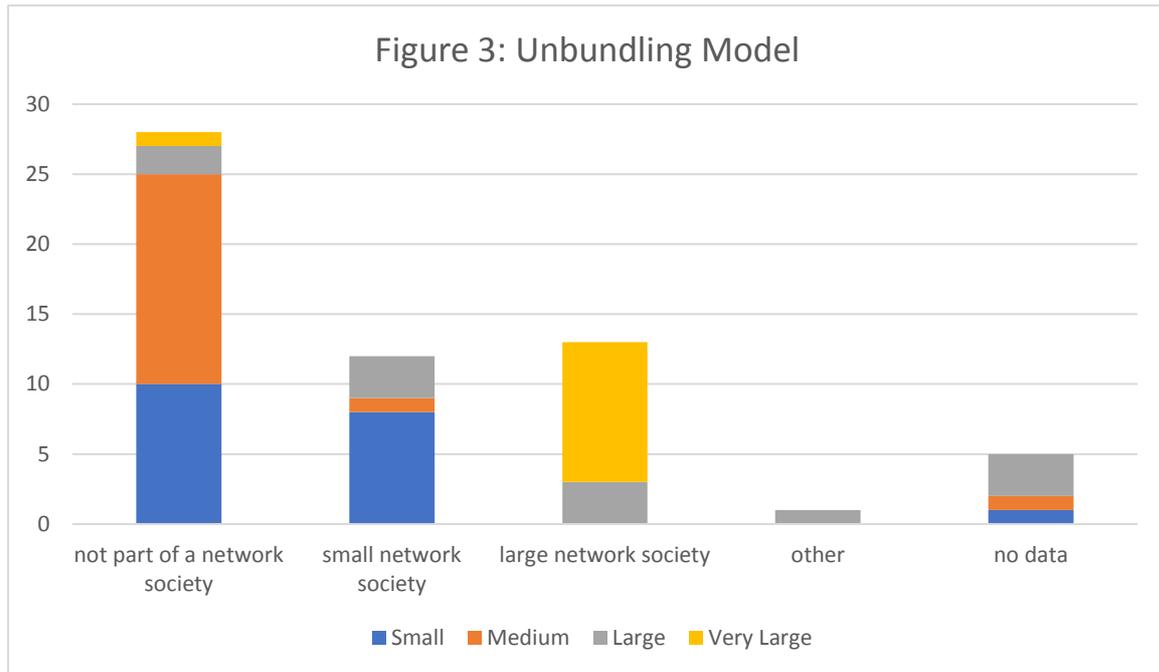

Figure 3: Which unbundling model is implemented in your company?

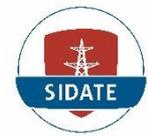

## Part B: Organisational Aspects

In this section of the survey, questions about organisational matters were posed. Subsequently, questions about the particularities of the responding employees were included. Examples thereof include items about their job position within the company, or the department they are part of. Furthermore, concrete questions about the IT security were asked.

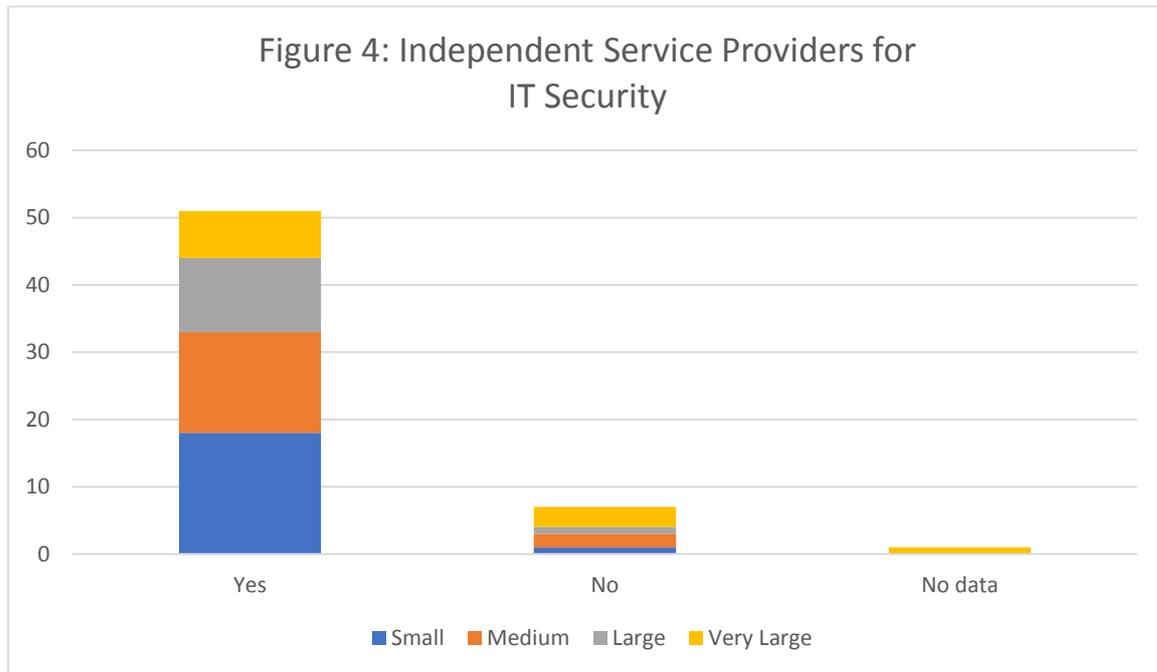

Figure 4: Are independent service providers in the field of IT security in your company?

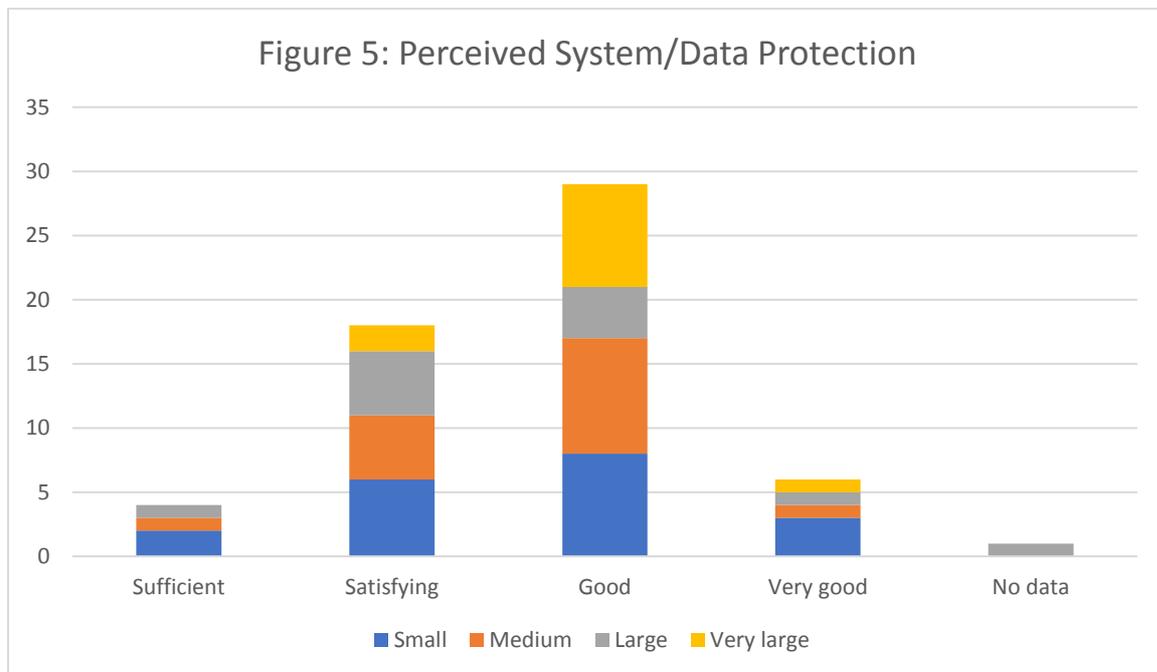

Figure 5: In your view, how well protected are the systems and data in your company?

# Part C: Information Security Management System (ISMS)

In order to provide an overview of the implementation status of information security management systems, the survey was designed with specific related questions.

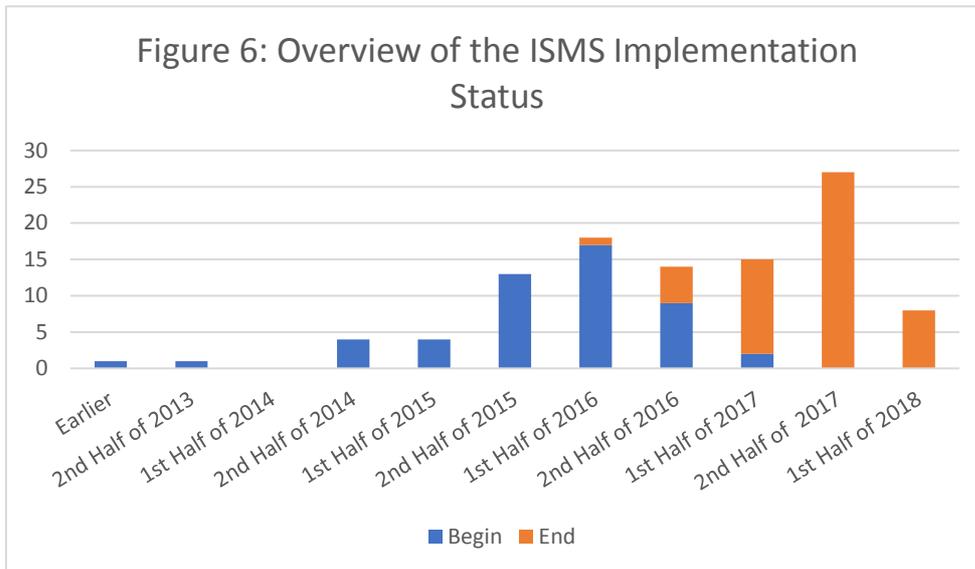

Figure 6: Overview of the ISMS Implementation Status

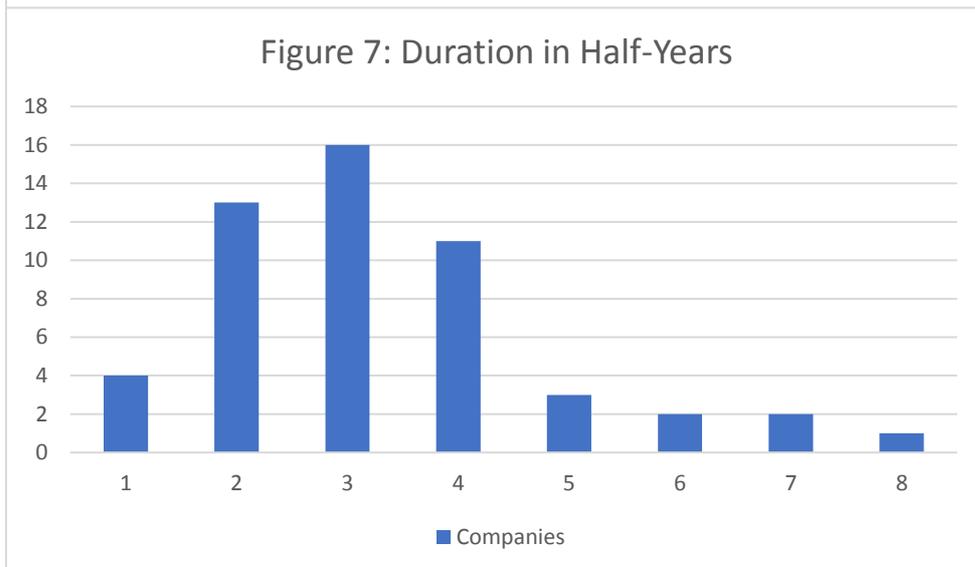

Figure 7: Duration in Half-Years

Figures 6 and 7 present the results of the following survey items:

- The implementation of ISMS …
- When are the ISMS implementation tasks supposed to start?
- When did the ISMS implementation begin?
- When is the ISMS implementation supposed to be finished?
- When was the ISMS implementation finished?

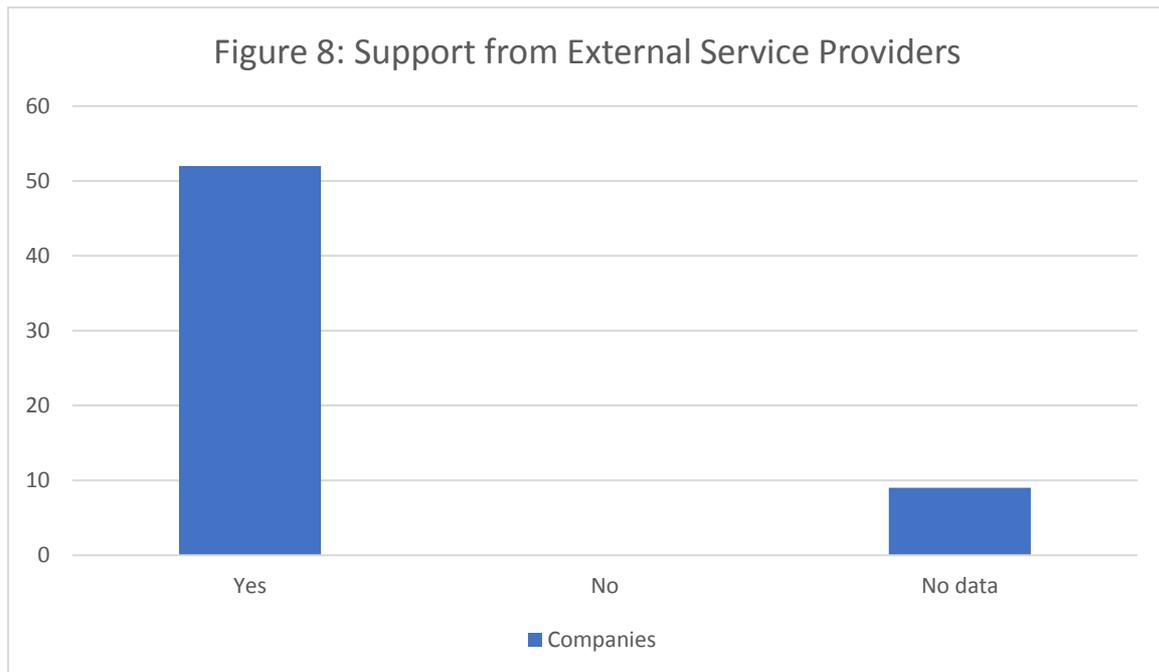

Figure 8: Were/Are external service providers (e.g. management consultants) subcontracted for the ISMS implementation?

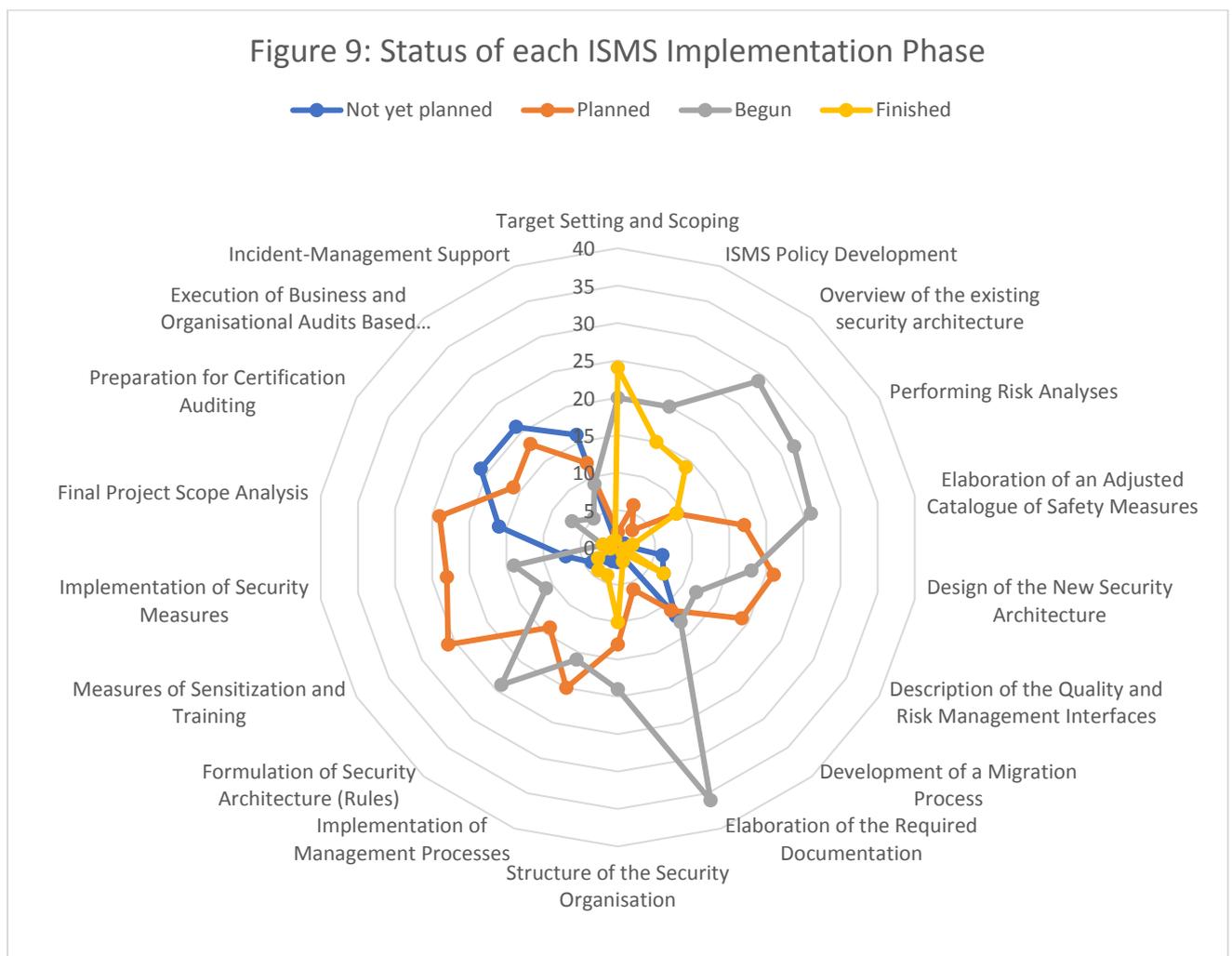

Figure 9: What is the current status of each ISMS implementation phase?

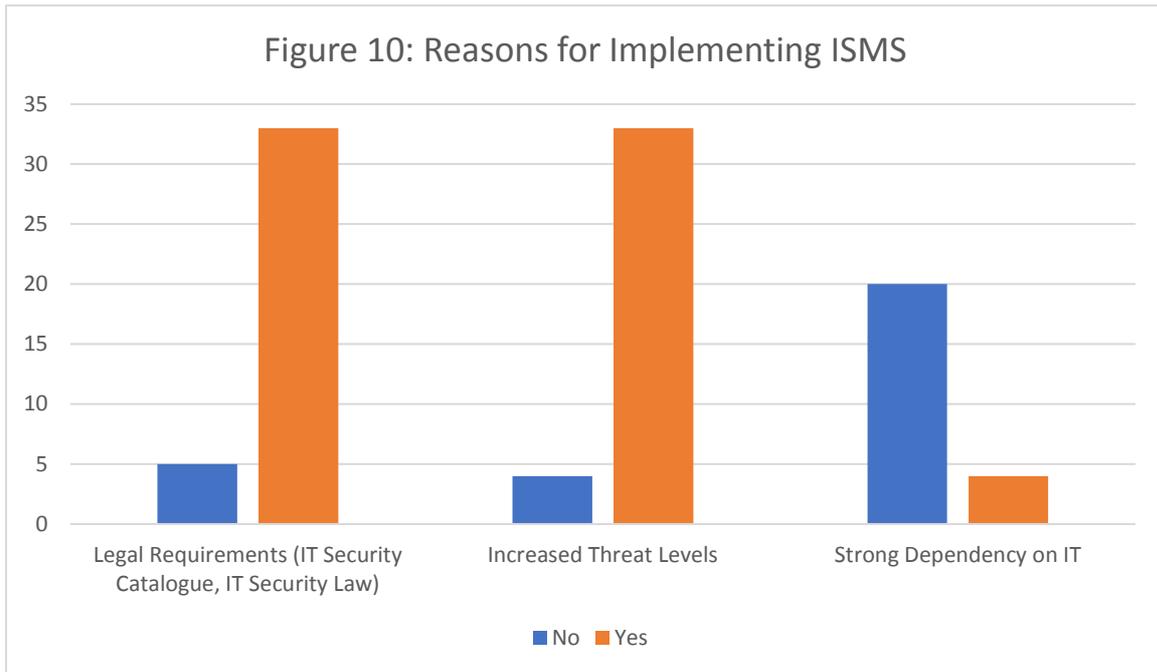

Figure 10: What were your reasons for implementing ISMS? (Multiple selection possible)

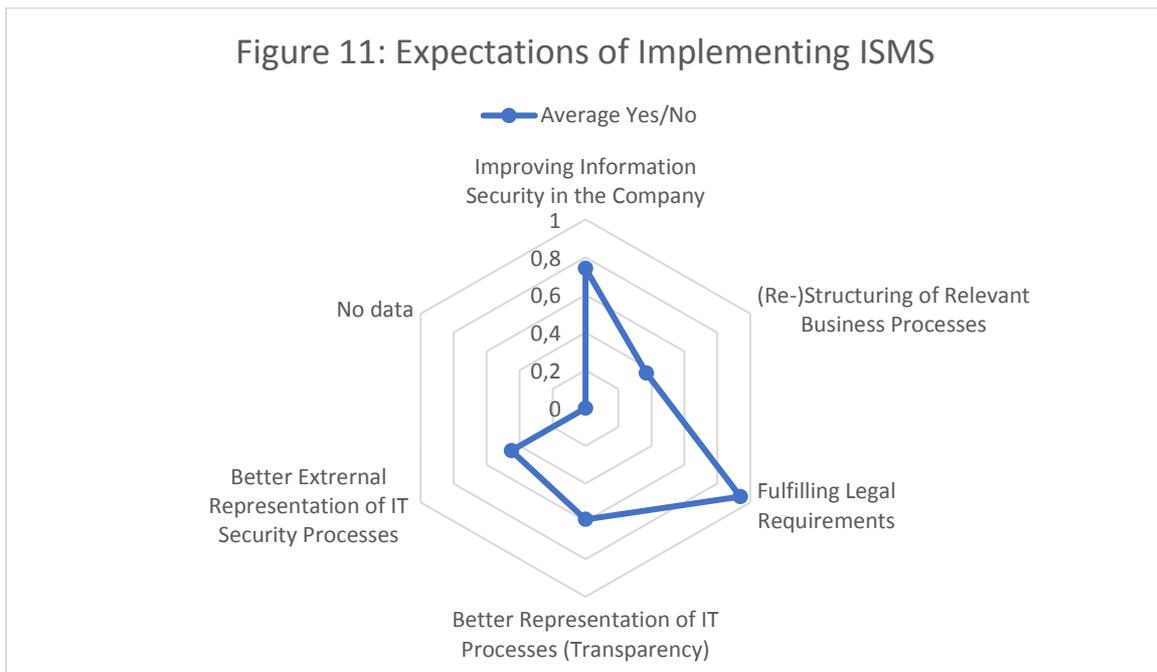

Figure 11: What are your hopes and expectations with regard to the ISMS implementation? (Multiple selection possible)

## Part D: IT Department

This section of the survey deals with the aspects of the IT security of the office IT. In order to be able to guarantee higher security levels, there must be corresponding IT security guidelines, which must be evaluated regularly. The evaluation is important, given the continuous technical advances.

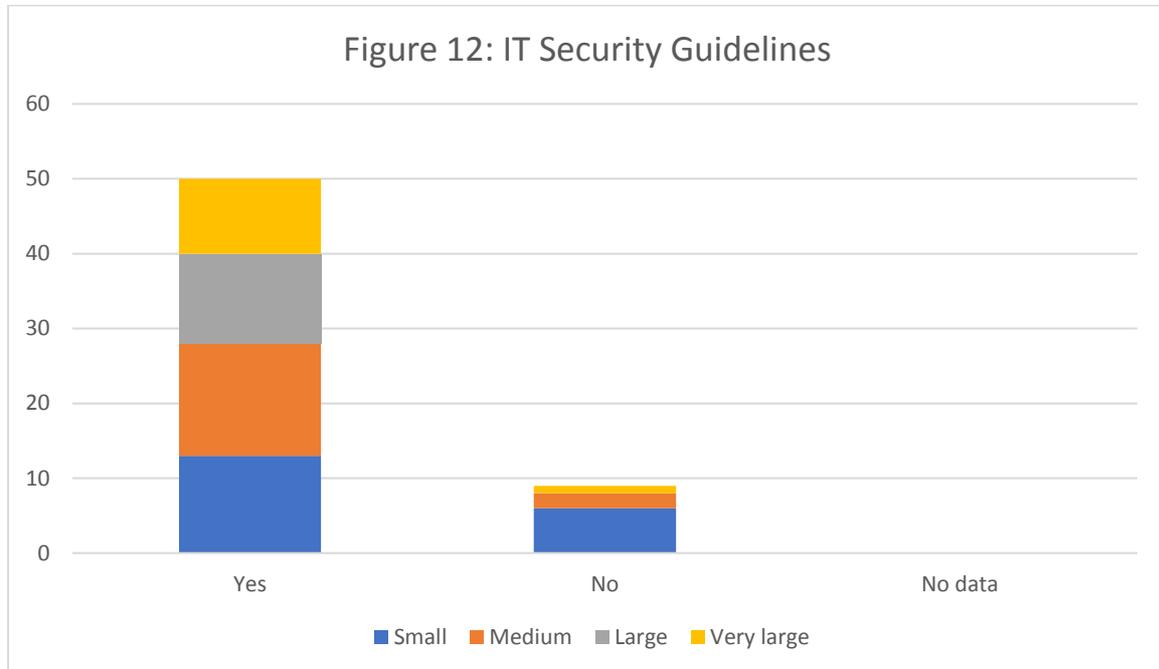

Figure 12: Are there IT security guidelines for the office IT in your company?

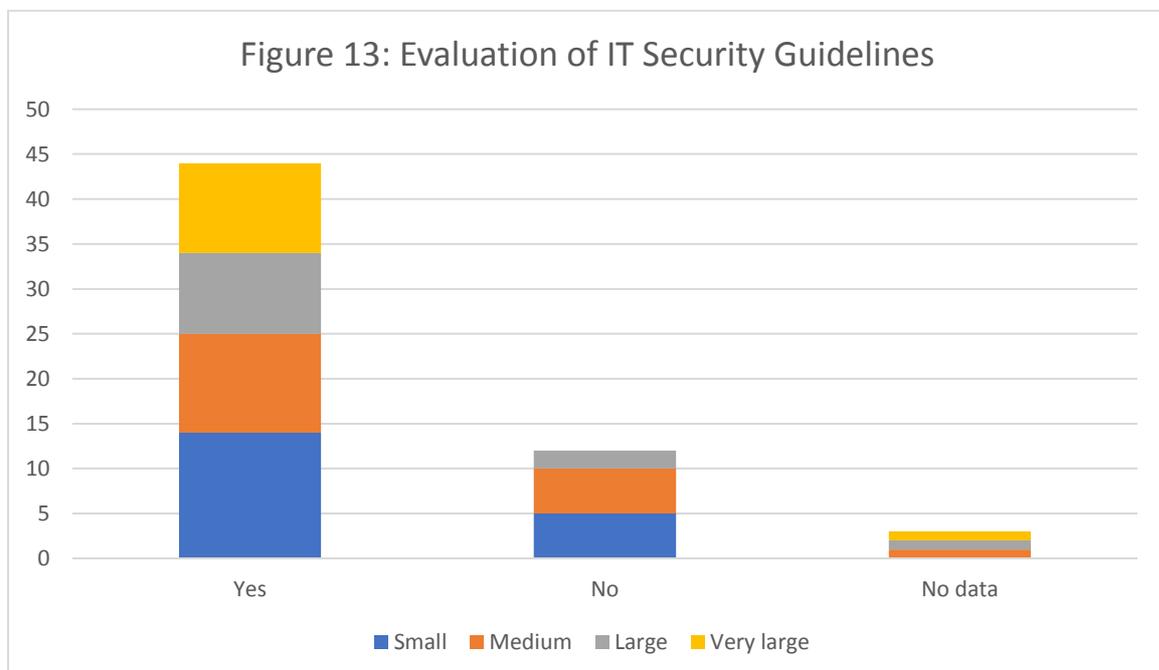

Figure 13: Are the IT security guidelines updated and, if necessary, adjusted regularly?

## Part E: Energy Control System: Network Structure

The energy control system is the core system of the energy providers. Particular questions about the network structure were included in the survey in order to gather more information on the general structure of the energy control system. Two main tasks of the energy control system are the energy network supervision and control, and the execution of switching operations.

Another important aspect of the network structure is the separation between the energy control system and other networks (e.g. office IT; Internet; 3rd parties). If there is no separation threats and potential attack points may exist and need to be mitigated.

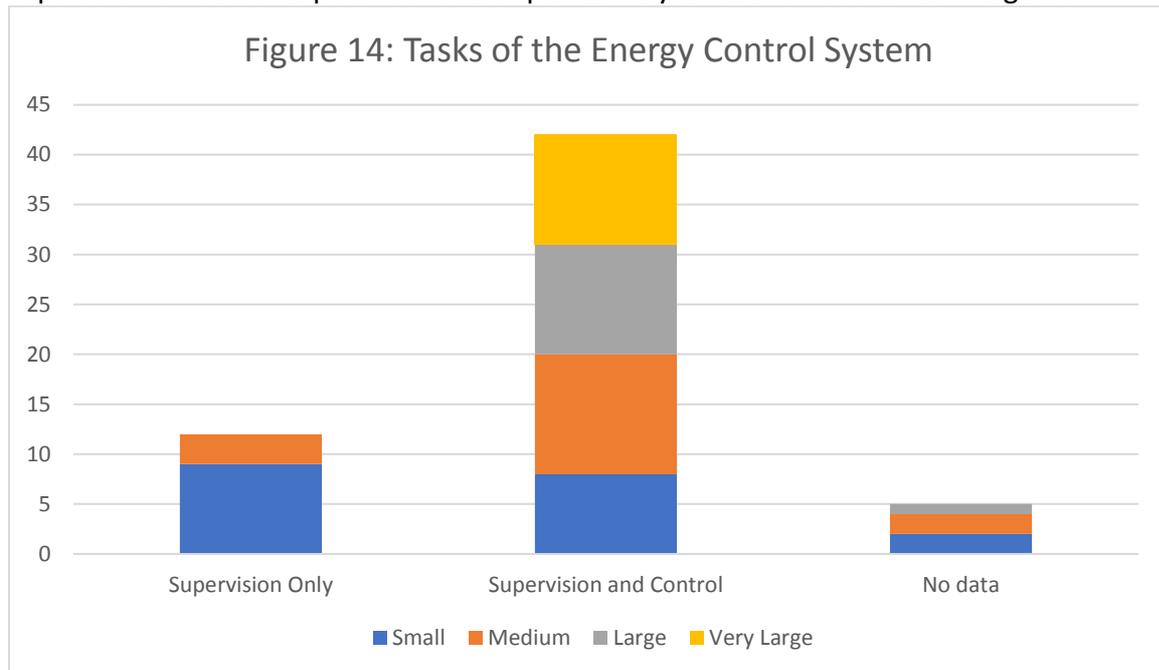

Figure 14: Does your energy control system undertake only energy network supervision, or can it also execute switching operations?

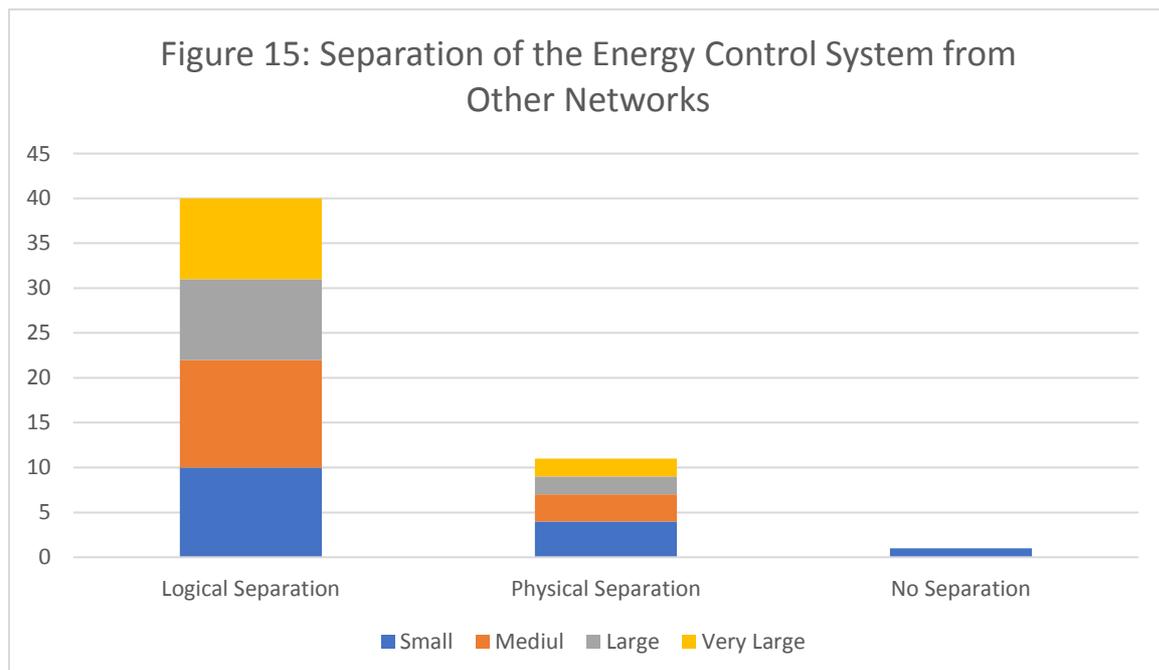

Figure 15: How is the IT network of your energy control system separated from other networks (e.g. IT department, Internet, maintenance companies)?

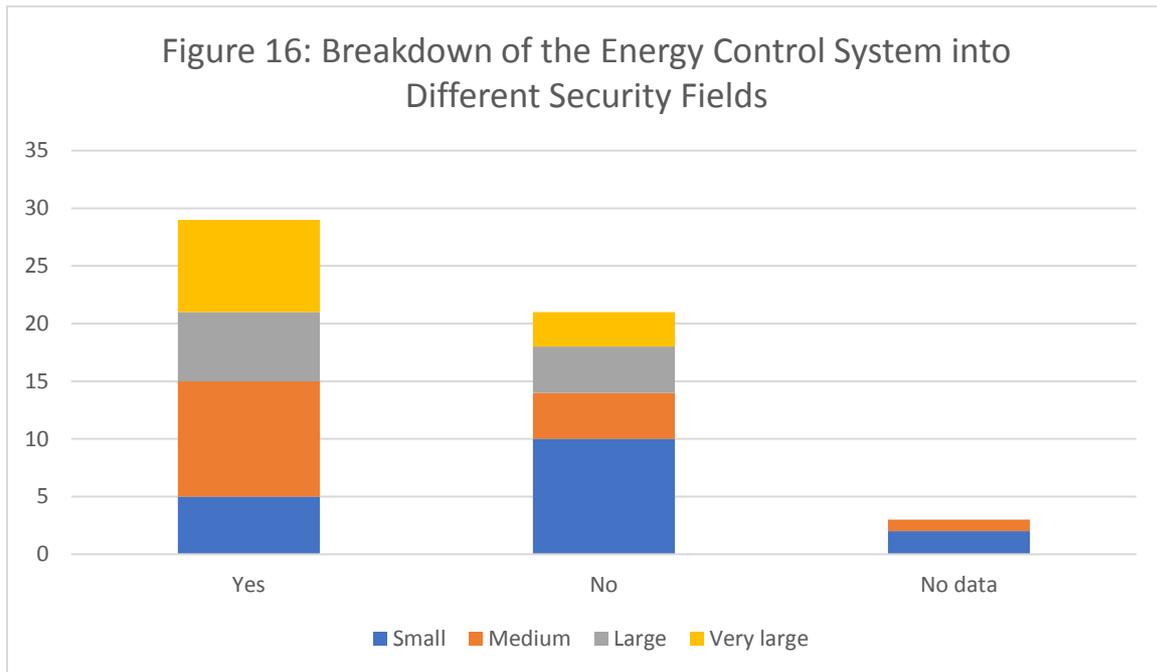

Figure 16: Is the network of your energy control system divided in different security domains (e.g. through different VLANs)?

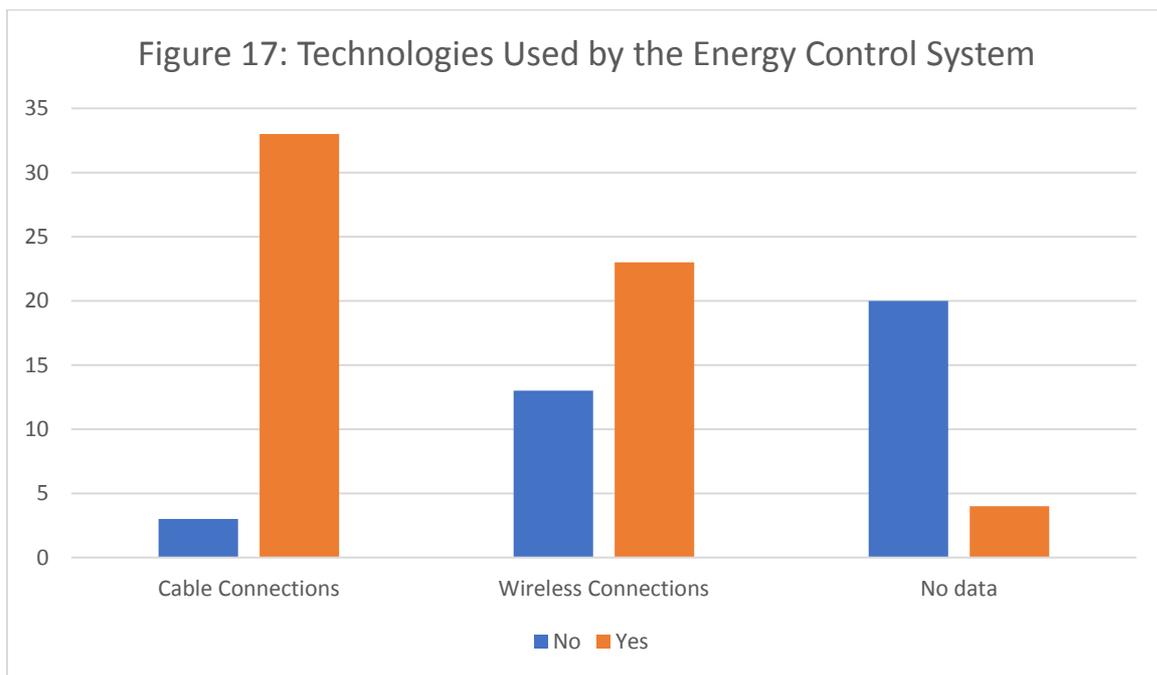

Figure 17: Which network technologies do you use in your energy control system network? (Multiple selection possible)

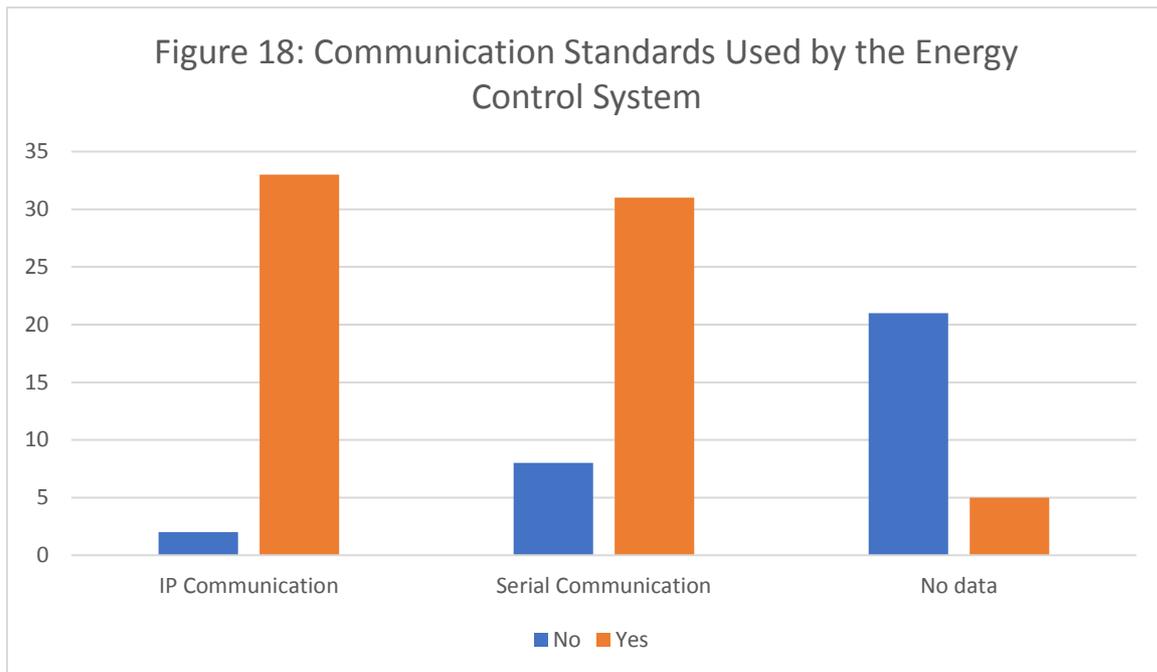

Figure 18: Which communication standards are used in the network of your energy control system?

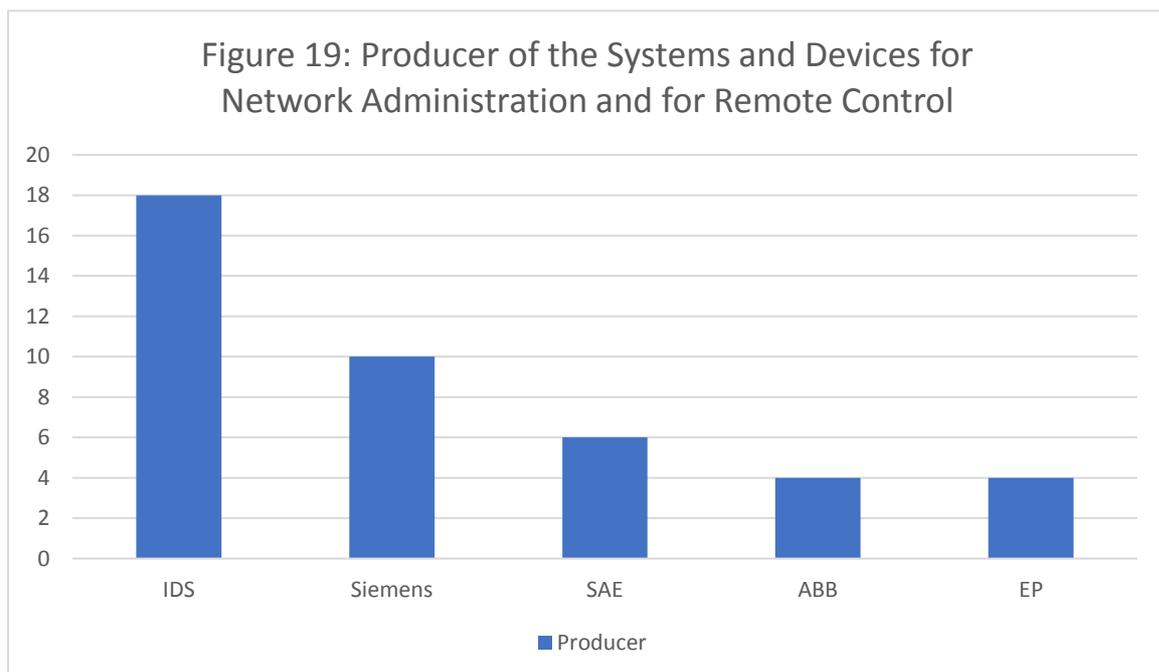

Figure 19: From which producers do you acquire the network administration systems and devices? (only producers with more than four nominations are depicted in the figure)

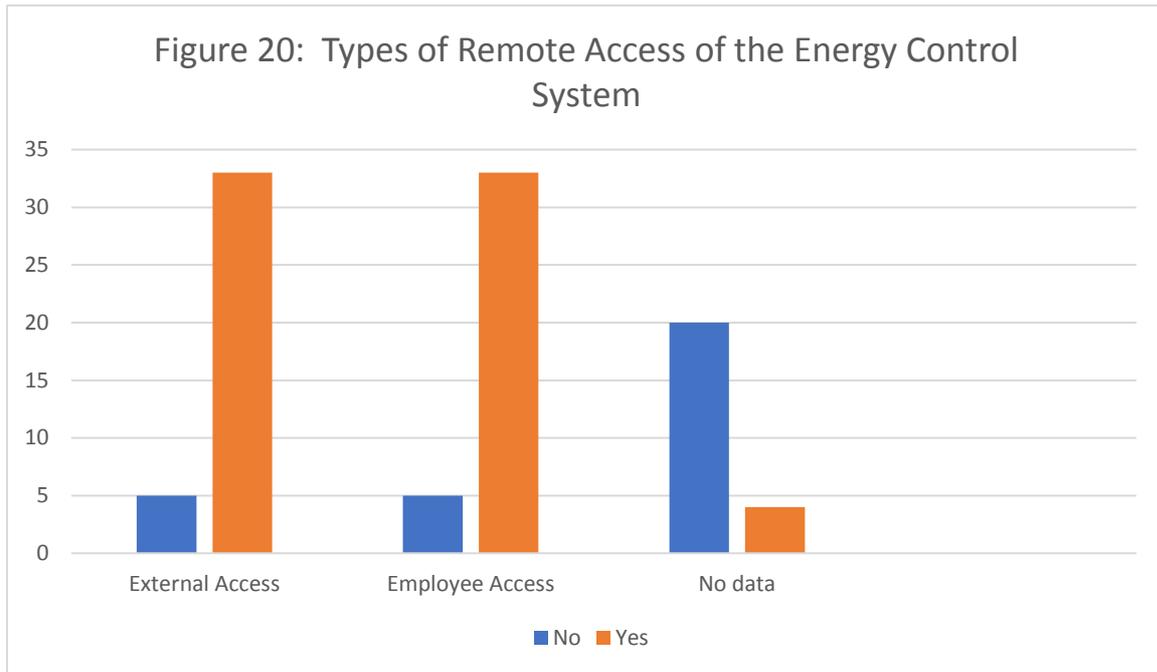

Figure 20: Which types of remote access were established for your energy control system? (multiple selection possible)

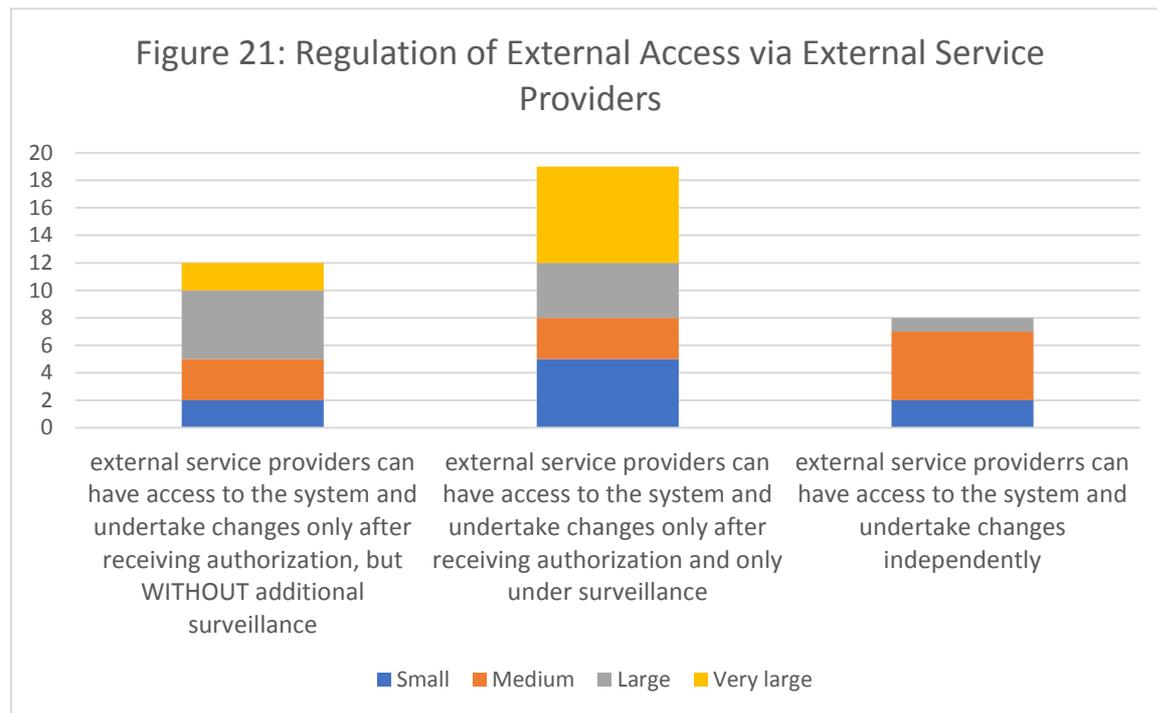

Figure 21: How are remote access procedures via external service providers regulated?

## Part F: Energy Control System: Processes and Organisation

Apart from the technical data of the energy control system, the underlying organisational structures and processes are also important. IT security must be continuously supervised and improved, since the means and technologies of potential attackers evolve constantly. At the same time, vulnerabilities must be dealt with, and there must be regular supervision and information dissemination within the company, such that hacker attacks can be prevented.

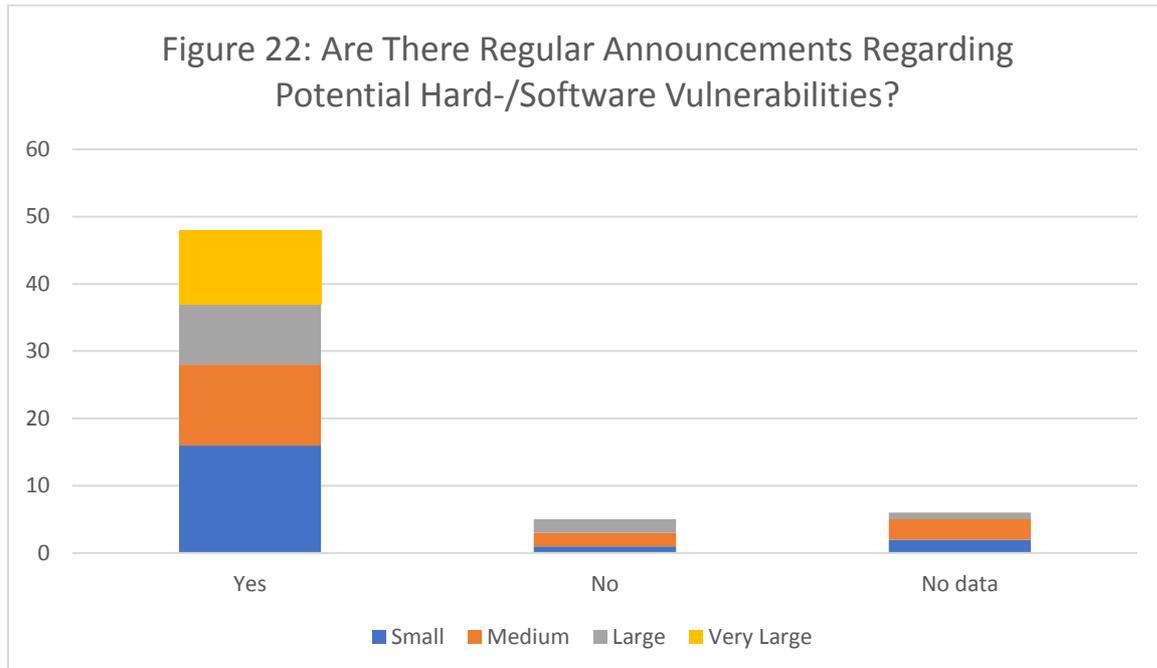

Figure 22: Are you/the responsible employees regularly informed about potential hard-/software vulnerabilities?

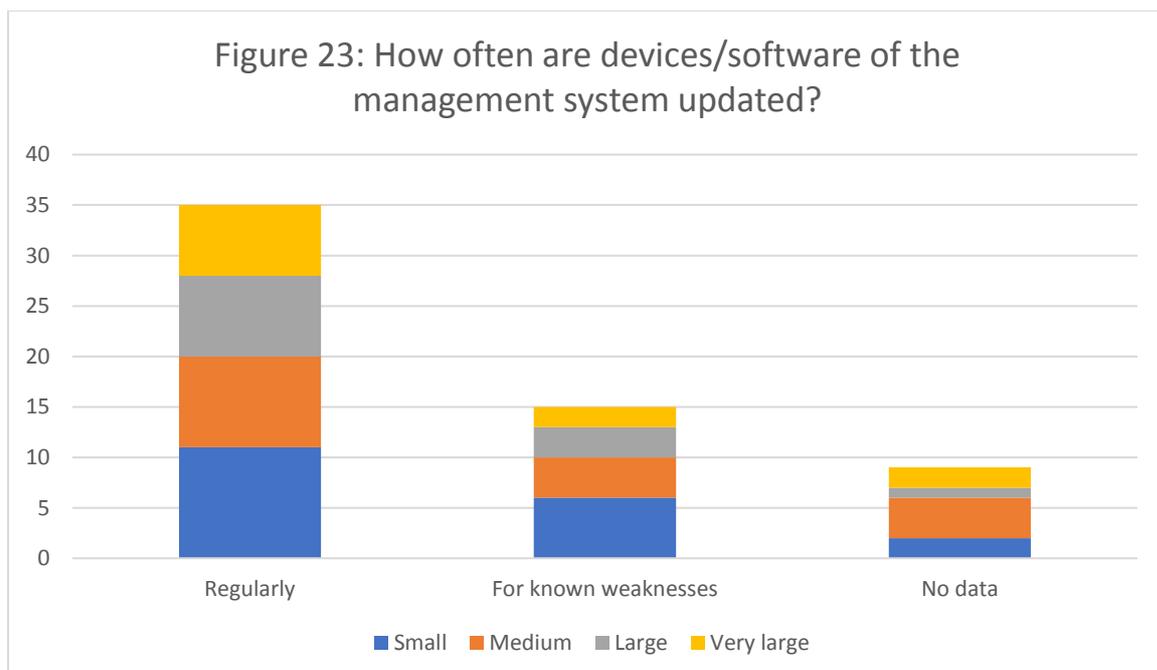

Figure 23: How often are the devices and software within your energy control system updated/renewed?

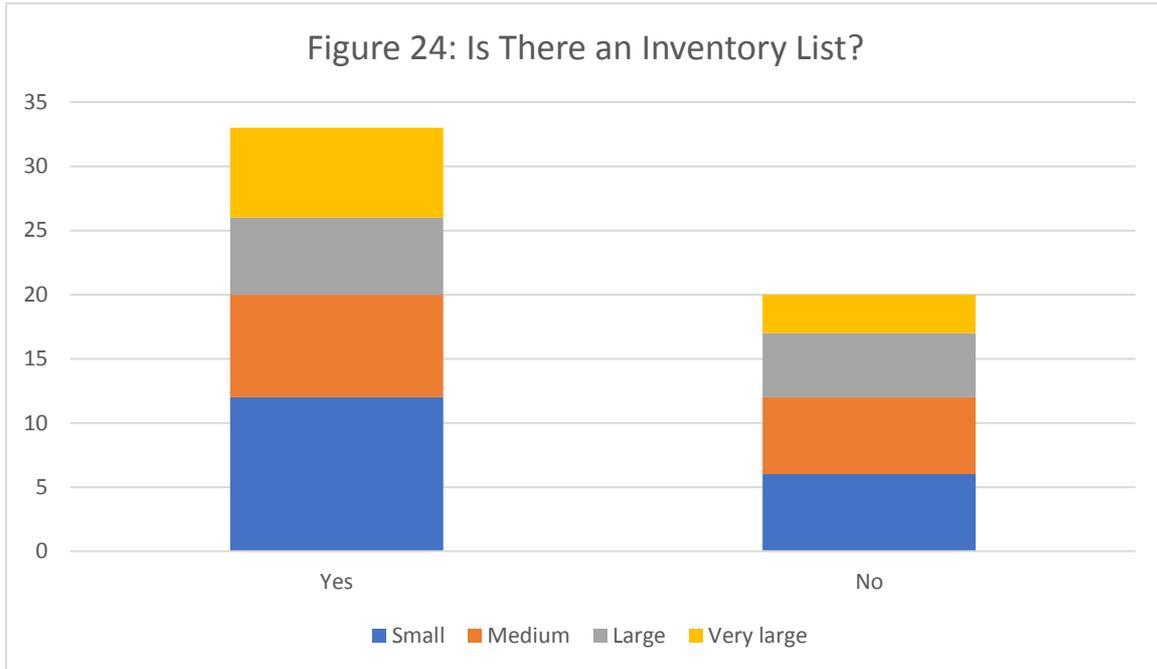

Figure 24: Is there an inventory list in which all the software items are documented (e.g. with version numbers, corresponding accounts and IP addresses)?

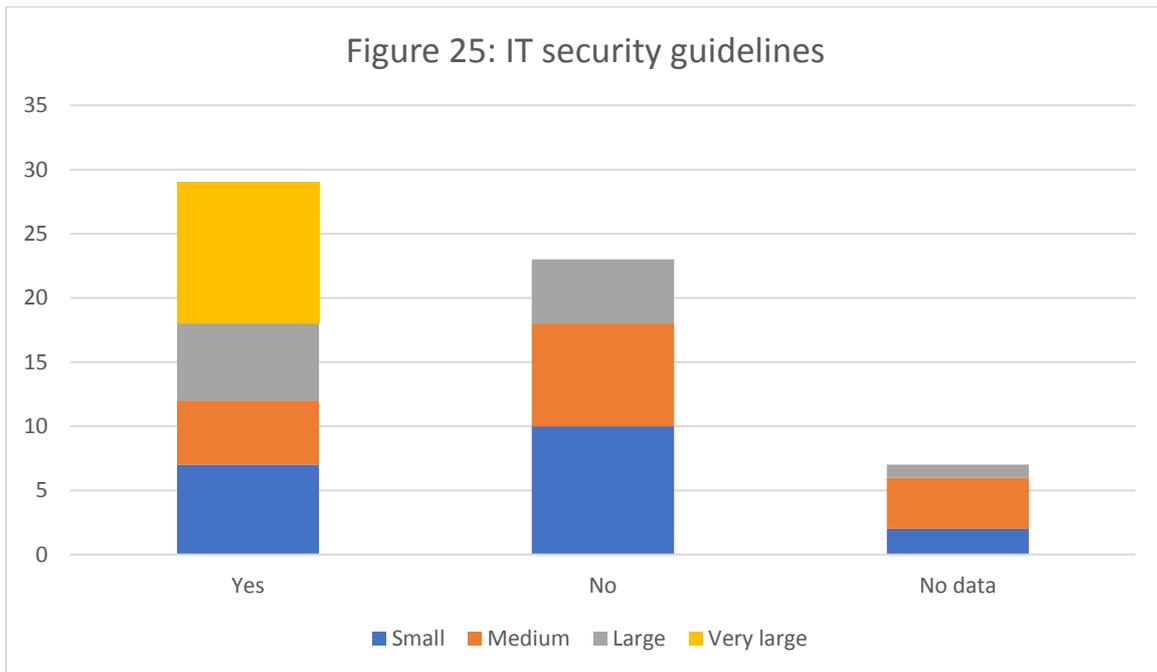

Figure 25: Are there recorded IT security guidelines for the energy control system in your company?

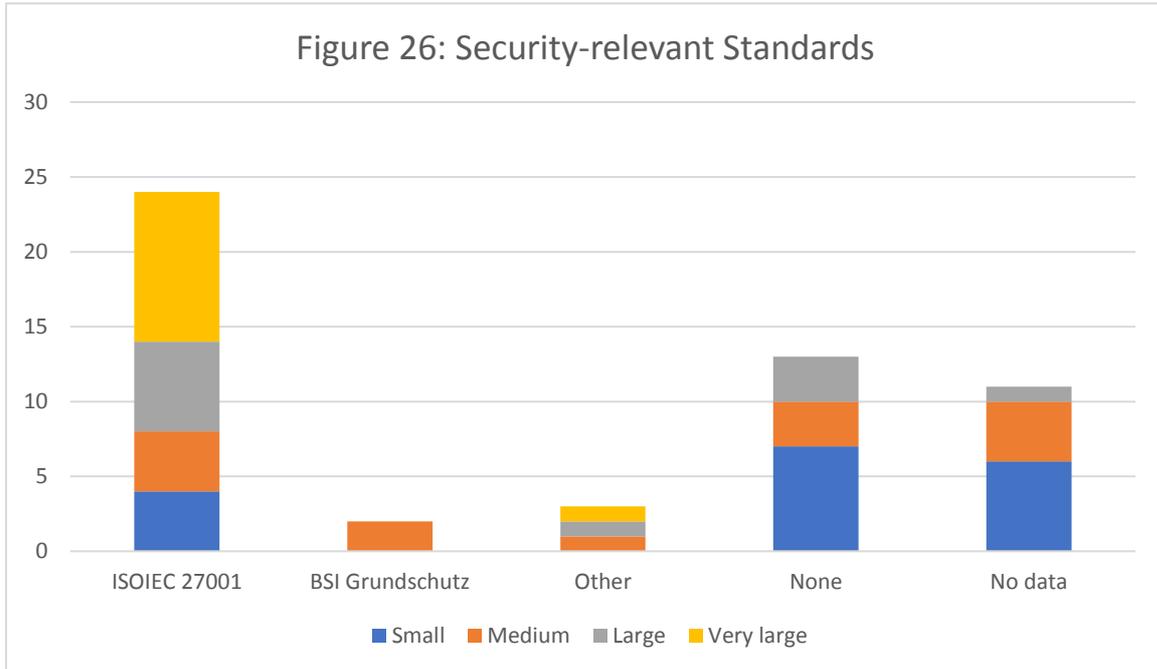

Figure 26: Under which security-relevant standards are your IT systems and processes for network administration elaborated?

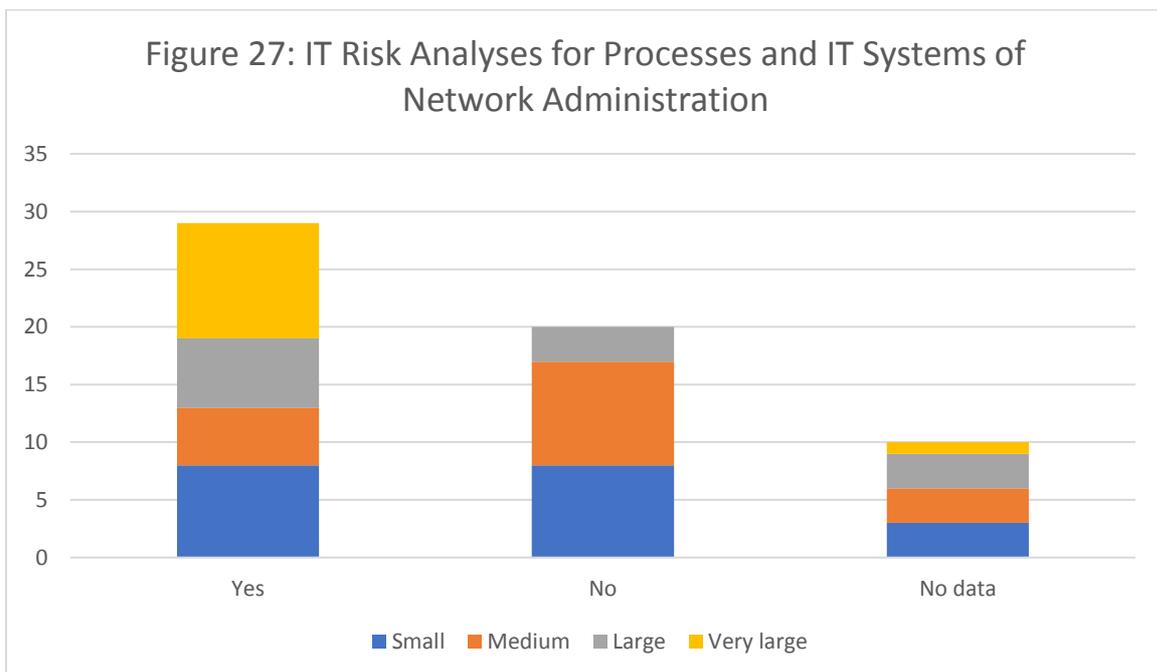

Figure 27: Do you perform IT risk analyses for the processes and IT systems for network administration?

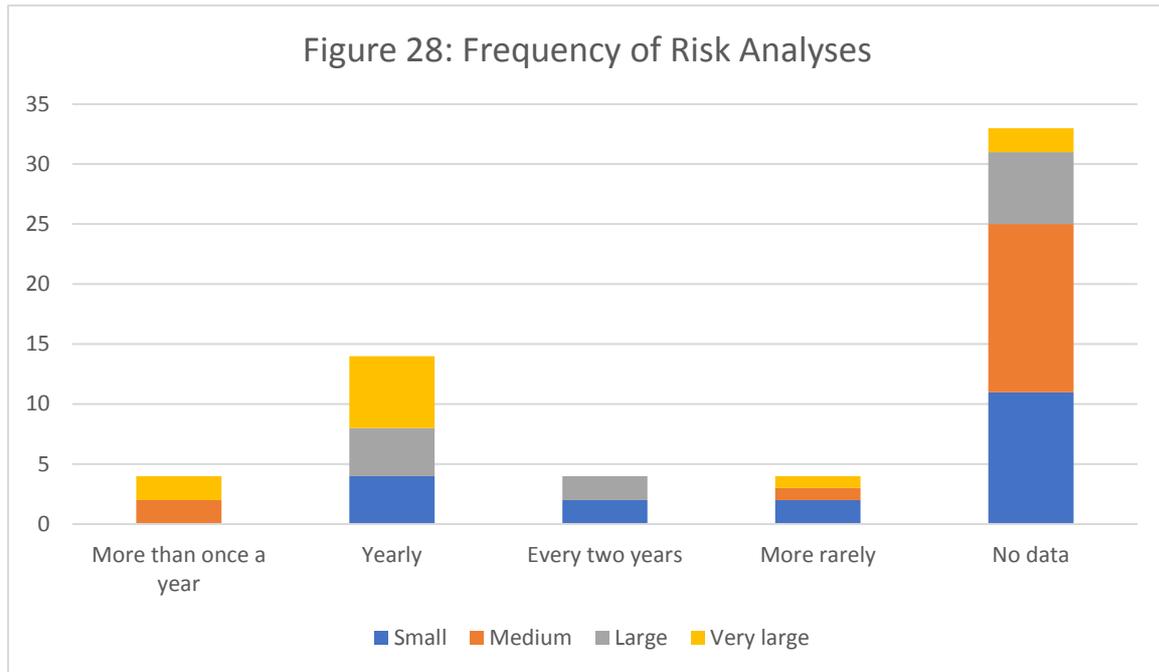

Figure 28: How often do you perform such risk analyses?

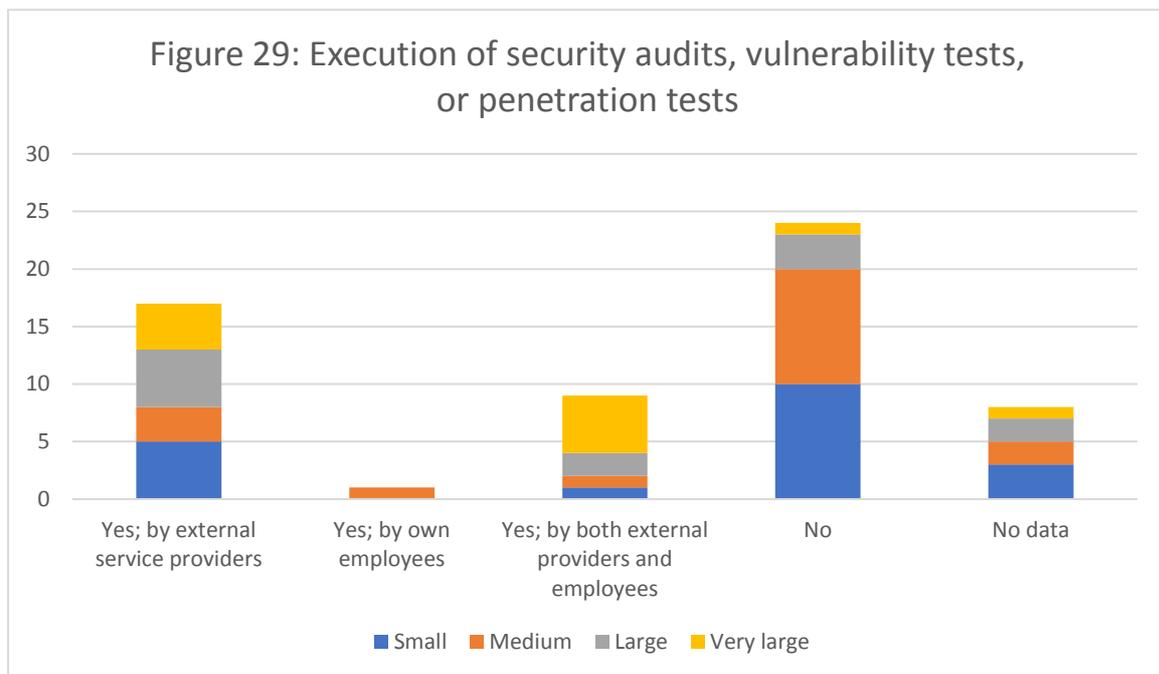

Figure 29: Do you perform security audits, vulnerability scans, or penetration tests for the administration systems of the network management technology?

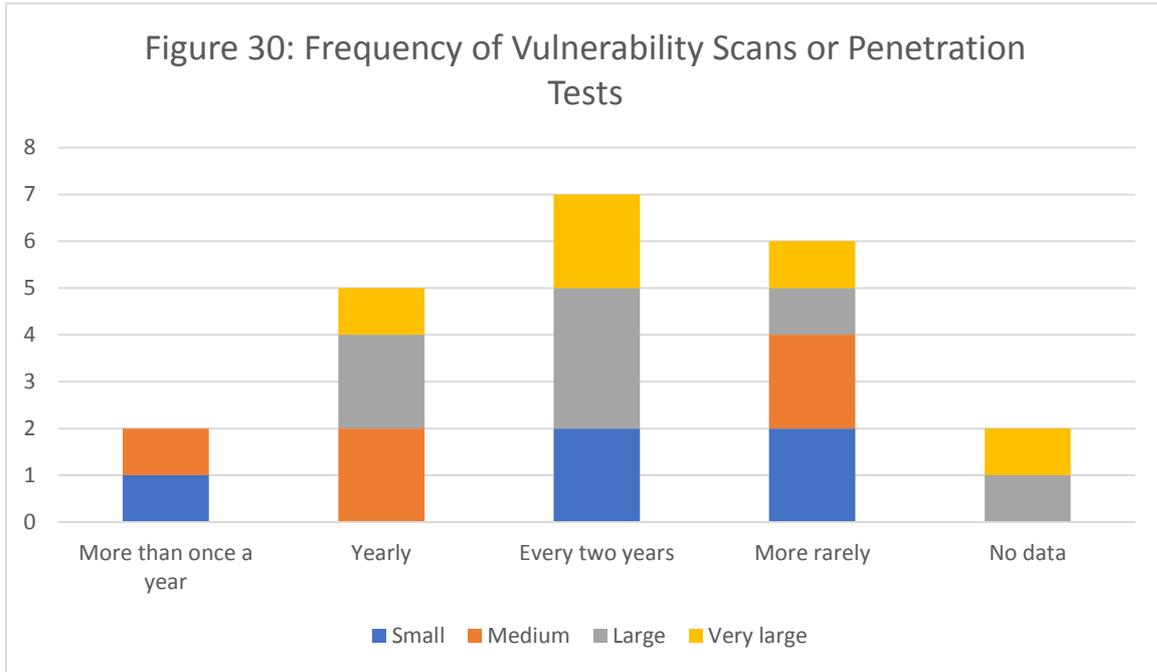

Figure 30: How often do you perform such vulnerability scans or penetration tests?

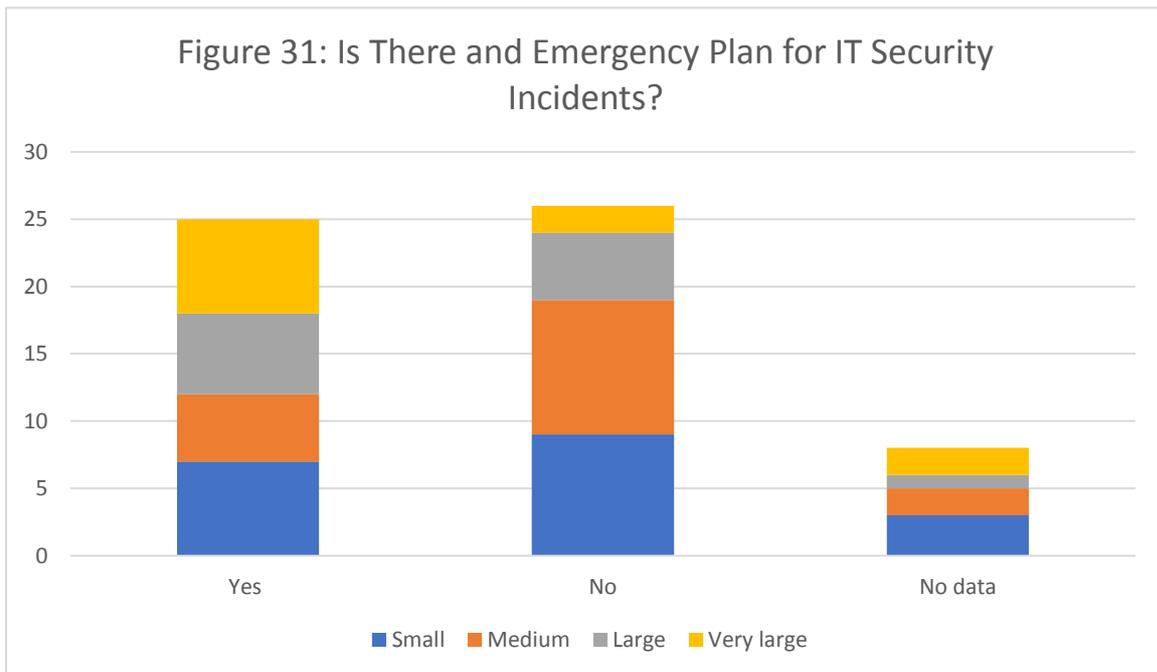

Figure 31: Do you have an emergency plan for security incidents of network administration?

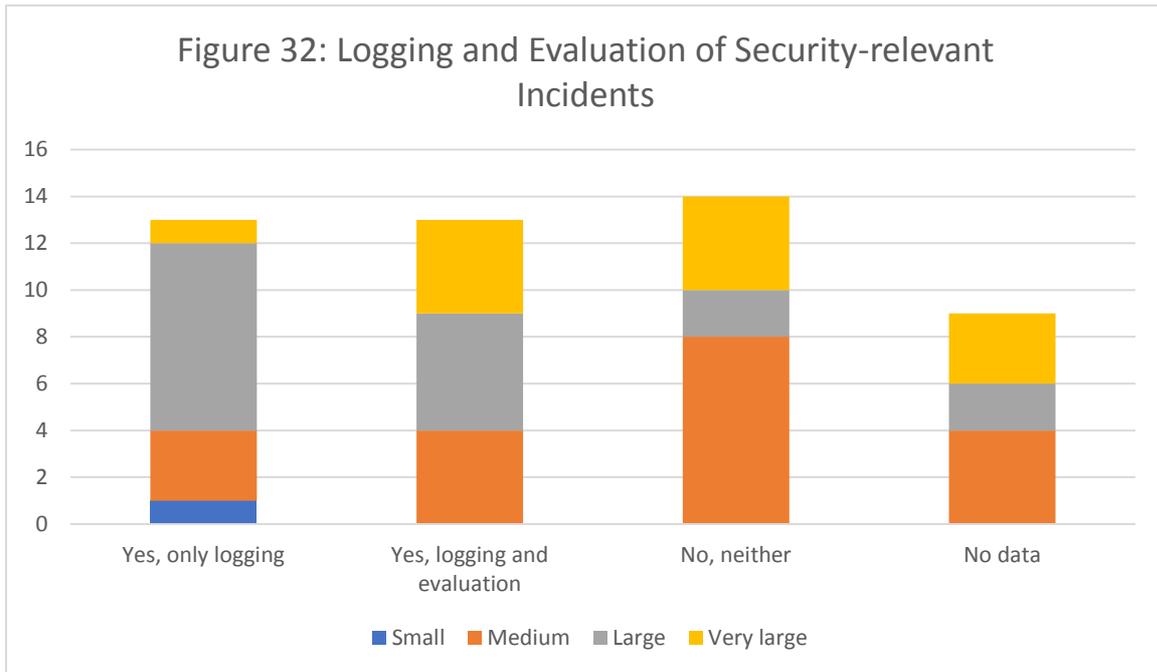

Figure 32: Are security-relevant incidents (e.g. portscans, failed login attempts, unauthorised processes) recorded and evaluated?

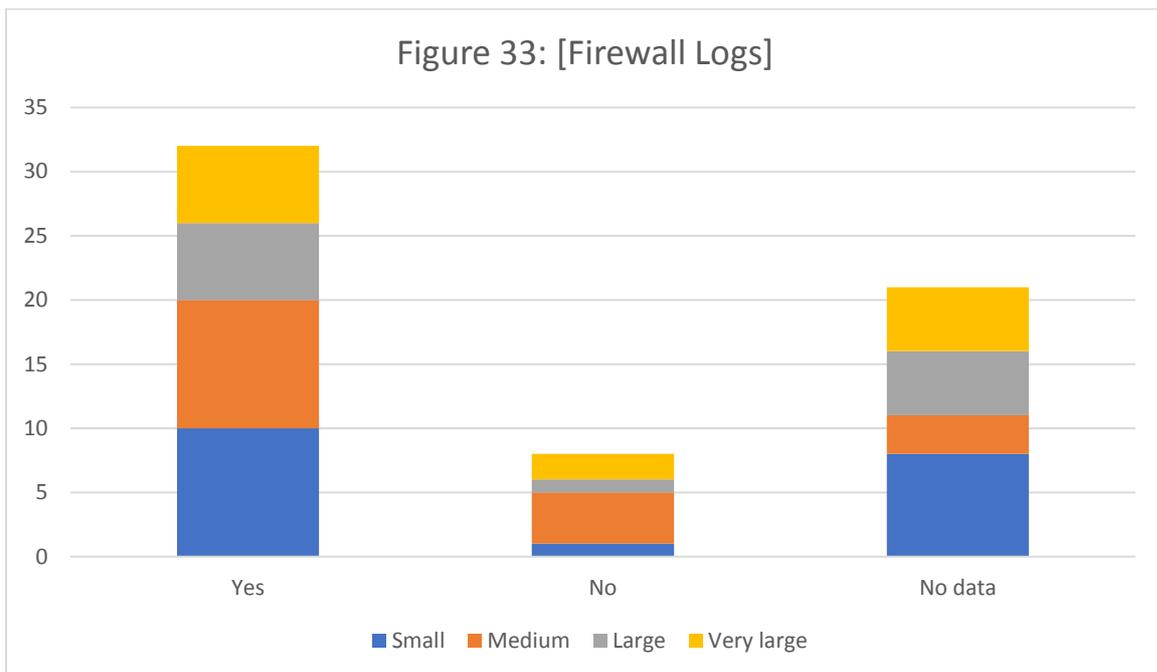

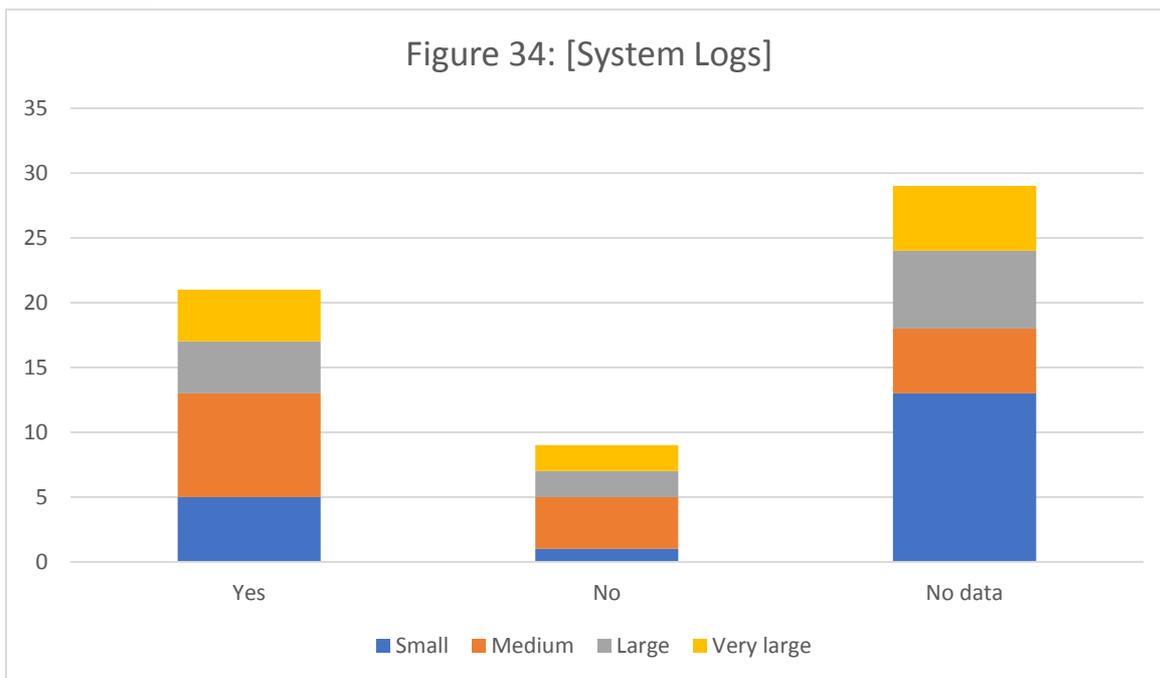

Figure 34: [System Logs]

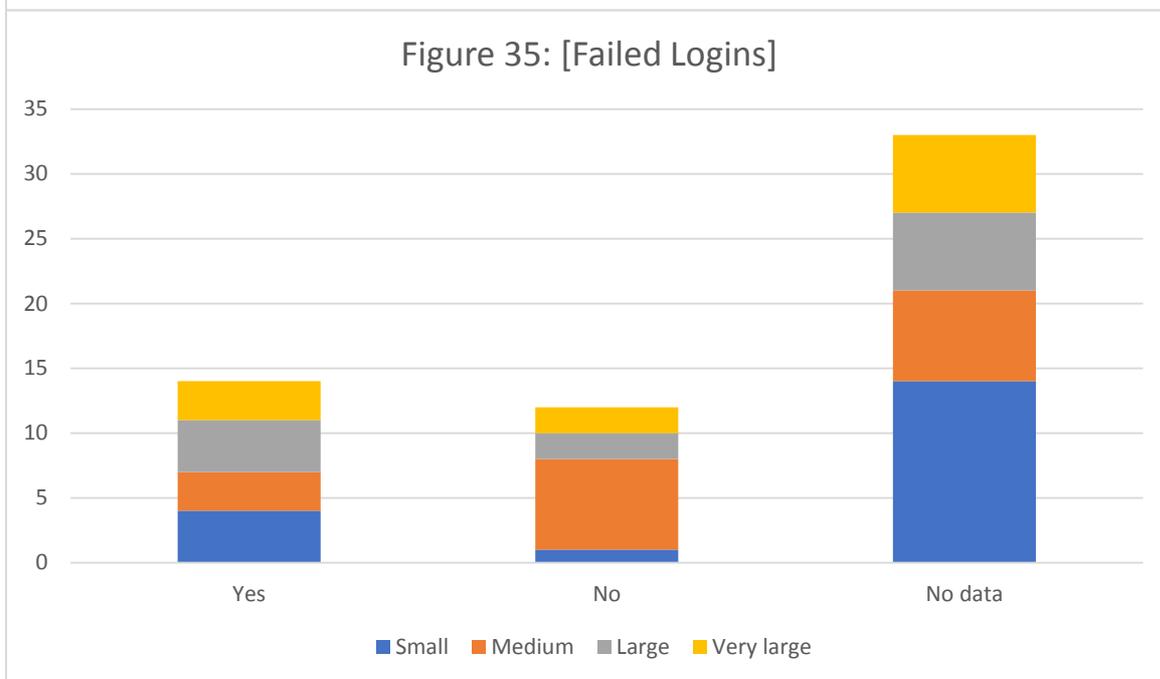

Figure 35: [Failed Logins]

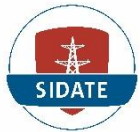

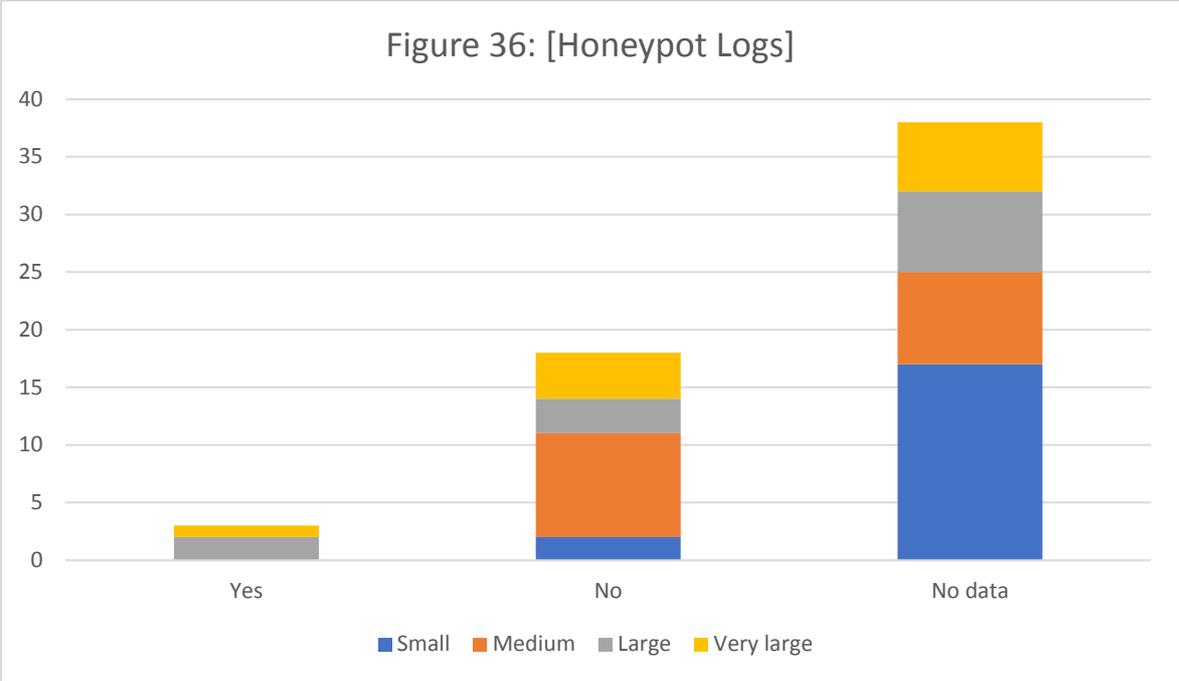

Figure 36: [Honeypot Logs]

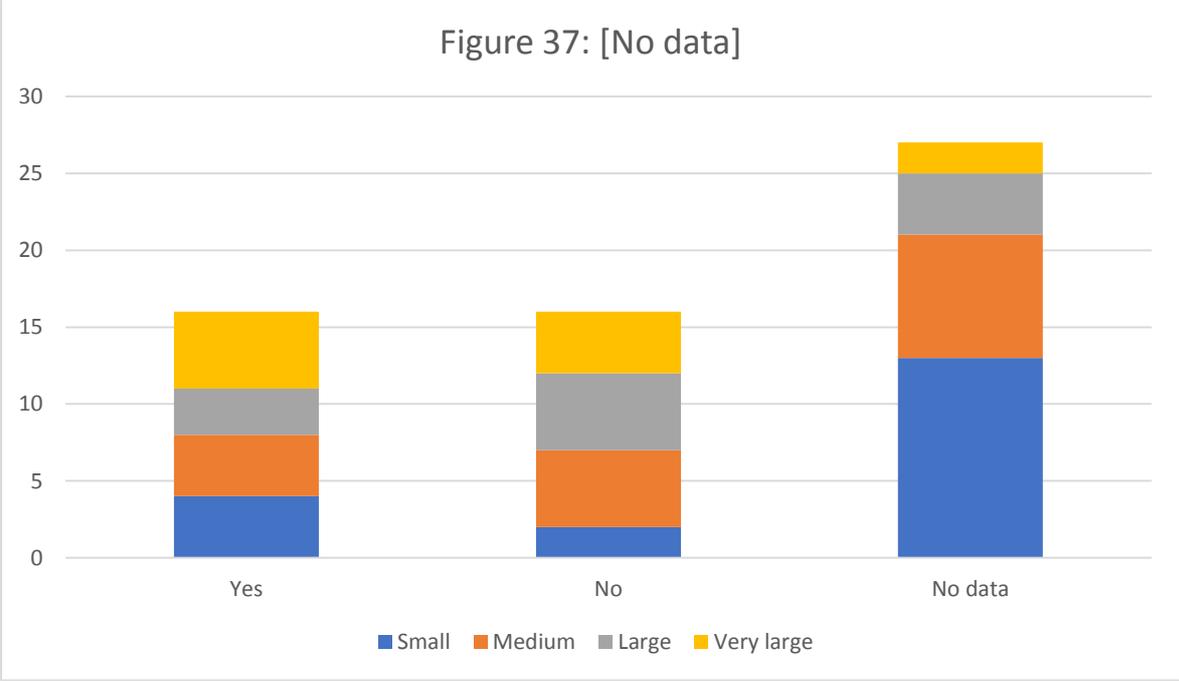

Figure 37: [No data]

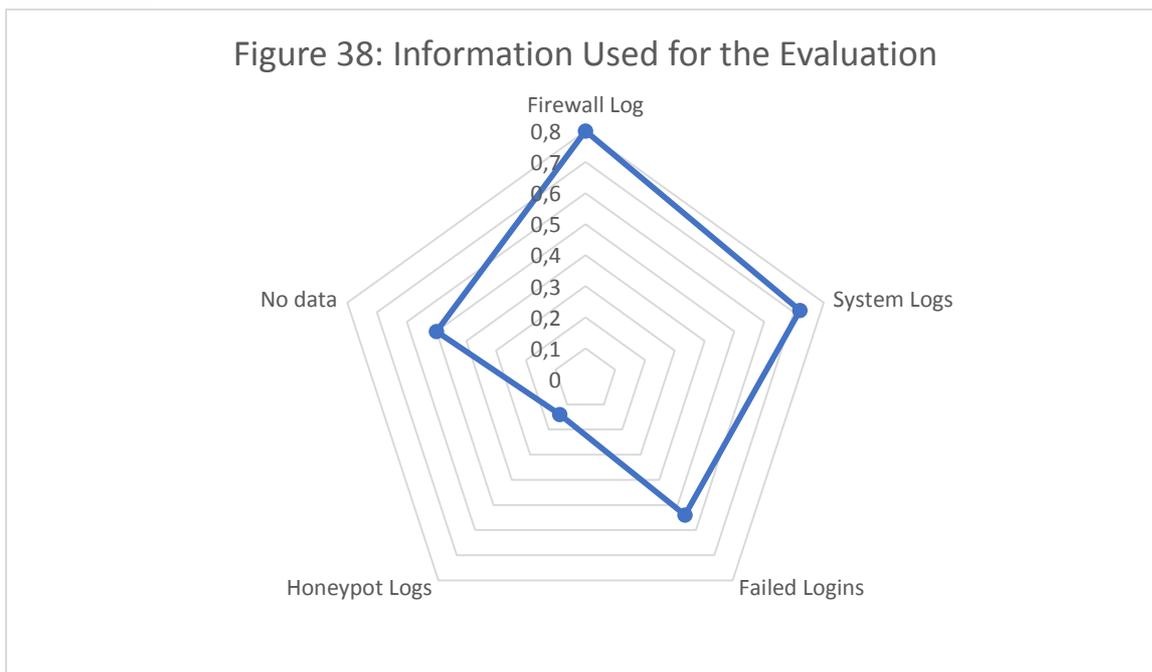

Figures 33 to 38: What information do you evaluate for identifying attacks in your IT systems for network administration? (Multiple selection possible)

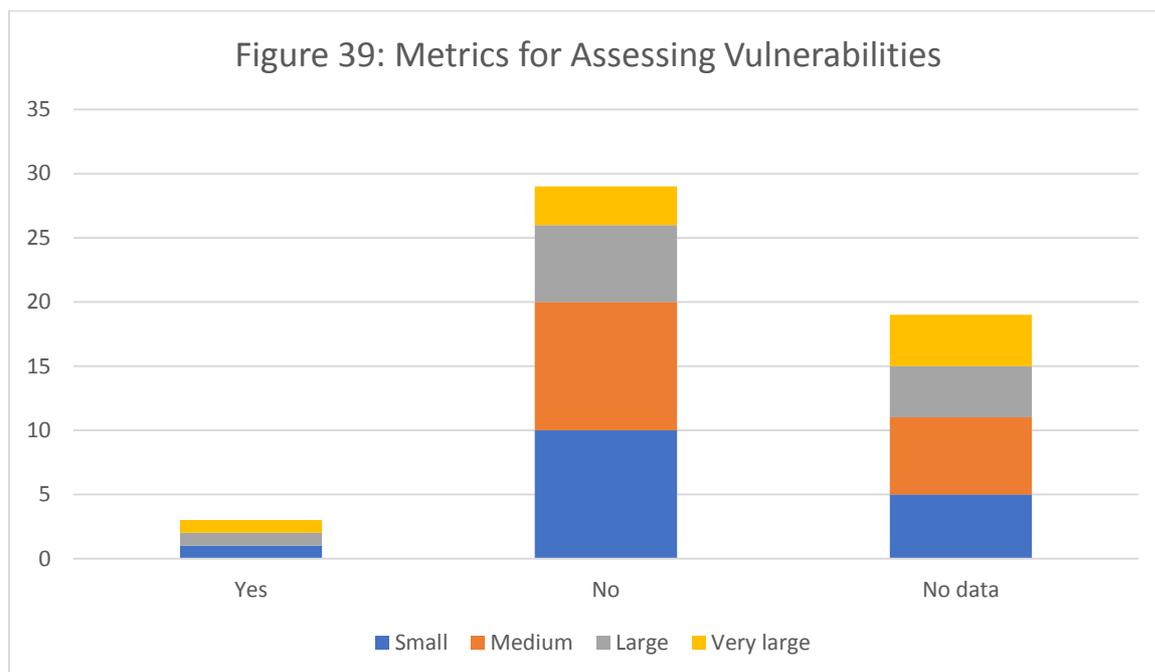

Figure 39: Do you use metrics to assess vulnerabilities (e.g. CVSS)?

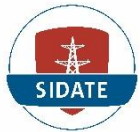

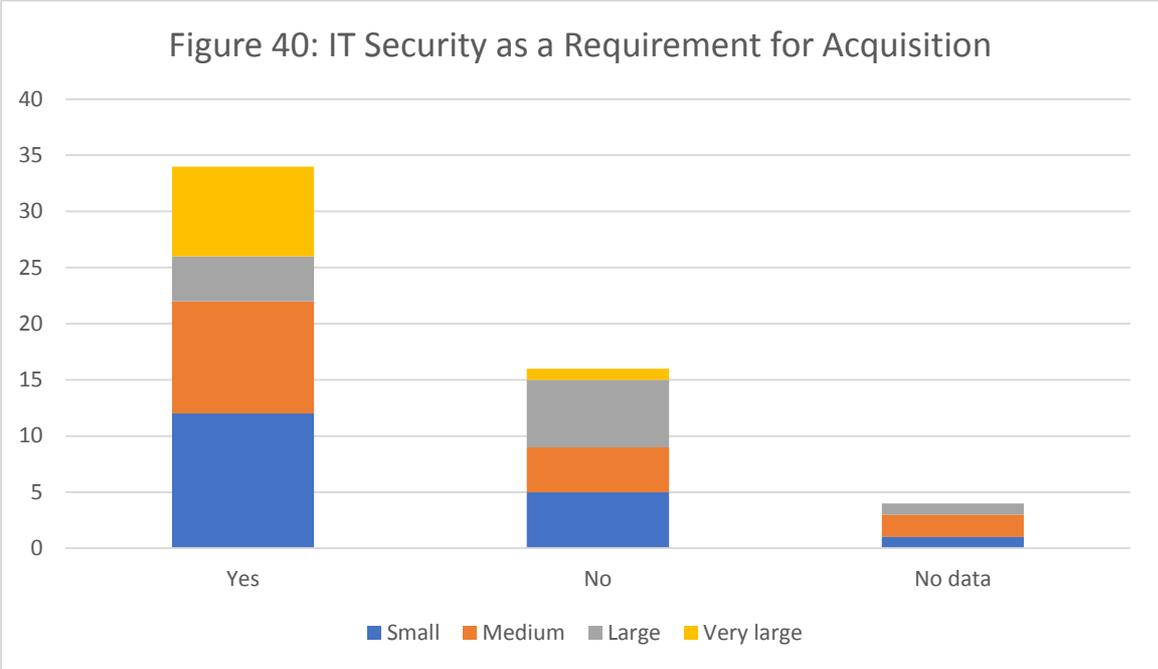

Figure 40: Is IT security defined as a requirement for acquiring new hard- and software?

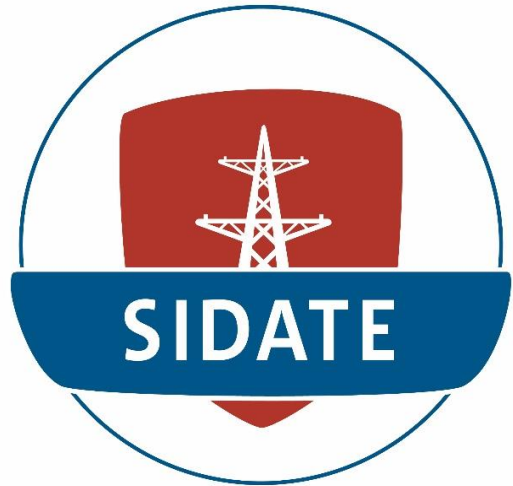

# Technischer Bericht

## Stand zur IT-Sicherheit deutscher Stromnetzbetreiber

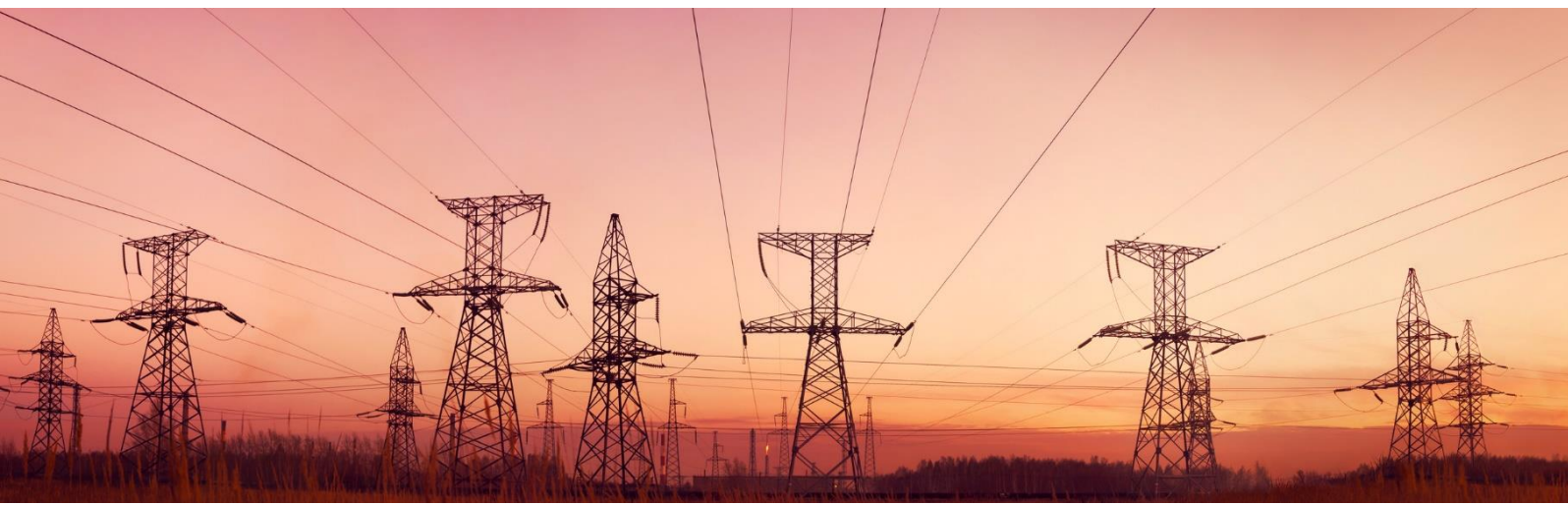

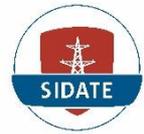

## Bei der Erstellung dieses Berichts haben mitgewirkt:

Julian Dax, Universität Siegen
Benedikt Ley, Universität Siegen
Sebastian Pape, Goethe Universität Frankfurt
Volkmar Pipek, , Universität Siegen
Kai Rannenberg, Goethe Universität Frankfurt
Christopher Schmitz, Goethe Universität Frankfurt
André Sekulla, Universität Siegen

## Sichere Informationsnetze bei kleinen und mittleren Energieversorgern (SIDATE)

Im Fokus des Forschungsprojekts SIDATE steht die technische Unterstützung kleiner und mittelgroßer Energieversorger bei der Selbsteinschätzung und Verbesserung ihrer IT-Sicherheit. Es werden verschiedene Konzepte und Werkzeuge in Zusammenarbeit von Universität Siegen, Goethe-Universität Frankfurt am Main, TÜV Rheinland i-sec GmbH, regio iT Gesellschaft für Informationstechnologie mbh, und der Arbeitsgemeinschaft für sparsame Energie- und Wasserverwendung (ASEW) entwickelt und evaluiert. Weitere Informationen finden sich auf der Webseite http://sidate.org/ .

### Förderhinweis

### Bildnachweis

Titelbild ©TebNad / Fotolia

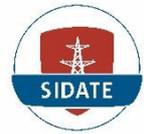

## Inhaltsverzeichnis

# Einleitung

Innerhalb des Forschungsprojektes „Sichere Informationsnetze bei kleinen und mittleren Energieversorgern" (SIDATE) wurde eine Umfrage zum Stand der IT-Sicherheit bei deutschen Stromnetzbetreibern durchgeführt. Das Projekt selbst beschäftigt sich mit der Informations-Sicherheit bei kleinen und mittleren Energieversorgern.

Zur Durchführung der Umfrage wurden alle 881 im August 2016 bei der Bundesnetzagentur gelisteten Betreiber angeschrieben. In dem Umfragezeitraum vom 1. September 2016 bis zum 15. Oktober 2016 antworten 61 (6.9%) der Betreiber. Der Fragebogen fokussiert die Umsetzung der rechtlichen Anforderungen und die Implementierung eines Informationssicherheitsmanagementsystems (ISMS). Weiterhin wurden Fragen zu dem Leitsystem, Netzaufbau, Prozessen, organisatorischen Strukturen und der Büro-IT gestellt. Nachfolgend werden alle auswertbaren Ergebnisse der Umfrage präsentiert. Einige Fragen wurden nur ungenügend beantwortet, sodass auf eine Präsentation dieser Ergebnisse verzichtet worden ist.

Die Umfrage gliedert sich in folgende Teilbereiche:
    A) Allgemeine Informationen zum Unternehmen
    B) Organisatorisches
    C) Information Security Management System (ISMS)
    D) Büro IT
    E) Leitsystem: Netzaufbau
    F) Leitsystem: Prozess und Organisation

Es gibt zwei unterschiedliche Arten von Balkendiagrammen. Die erste besitzt farblich nur blaue Balken. Diese beziehen sich auf alle Stromnetzbetreiber, welche die entsprechende Frage beantwortet haben. Hingegen gibt es bei der zweiten Art der Balkendiagramme eine kategorische Unterscheidung zwischen den Stromnetzbetreibern. Diese liegt in der Größe, welche anhand der Anzahl der jeweils zugehörigen Zählpunkte festgemacht worden ist.

In einigen Ausnahmen wurde eine Art des Spinnennetzdiagramms verwendet. Auch in diesen Fällen wurde auf eine Kategorisierung der antwortgebenden Unternehmen verzichtet.

## Teil A: Allgemeine Informationen zum Unternehmen

Um einen ersten Überblick zu den teilnehmenden Stromnetzbetreibern zu erhalten, wurden im ersten Abschnitt der Umfrage allgemeine Fragen gestellt. Anhand der erhaltenen Ergebnisse konnten die Stromnetzbetreiber in vier Größenkategorien unterteilt werden, um die darauffolgenden Fragen besser auszuwerten.

Die Kategorisierung der Teilnehmer ist anhand der Anzahl der Zählpunkte des Stromnetzbetreibers getätigt worden. In Abbildung 1 ist die Aufteilung der Größe gut erkennbar. Für die weiteren Ergebnisse sind die Teilnehmer in die Kategorien „klein" (0 bis 15.000 Zählpunkte), „mittel" (15.001 bis 30.000 Zählpunkte), „groß" (30.001 bis 100.000 Zähpunkte) und „sehr groß" (ab 100.001 Zählpunkte) eingeteilt.

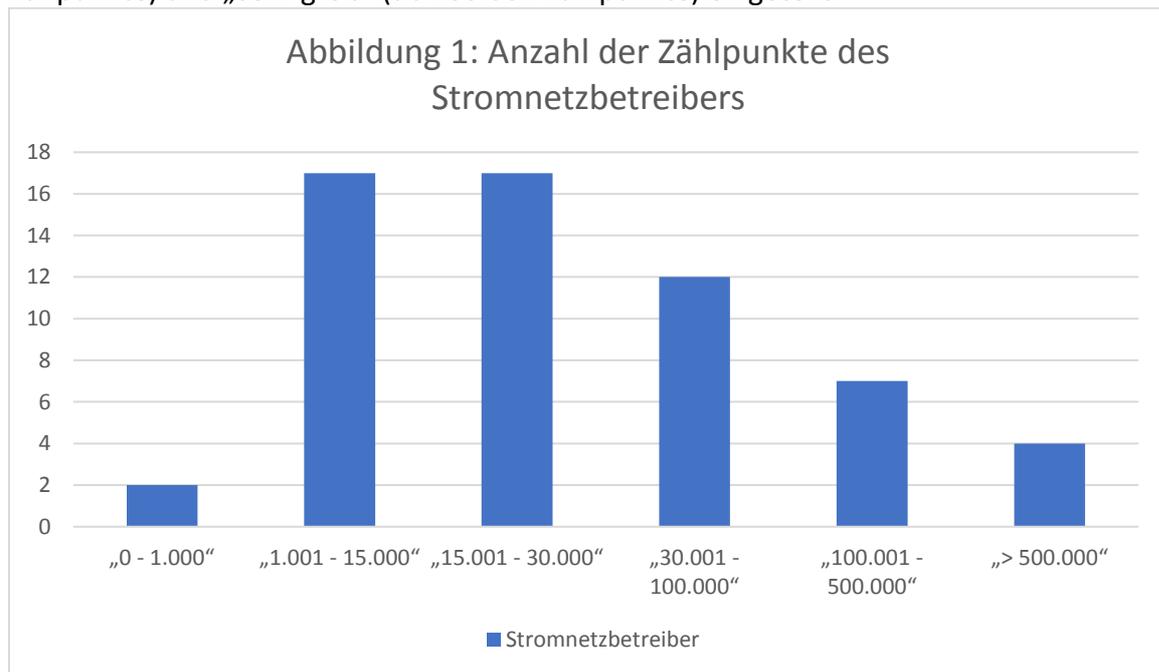

Abbildung 1: Wie viele Zählpunkte werden über Ihr Stromnetz versorgt?

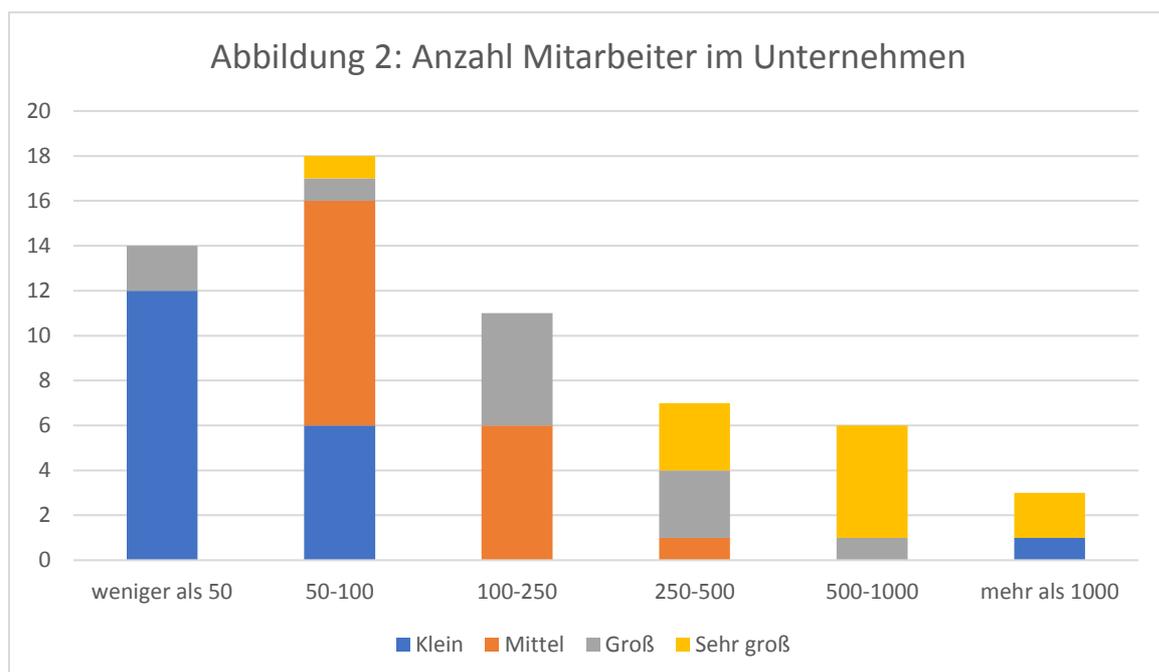

Abbildung 2: Wie viele Mitarbeiter/innen sind in Ihrem Unternehmen beschäftigt?

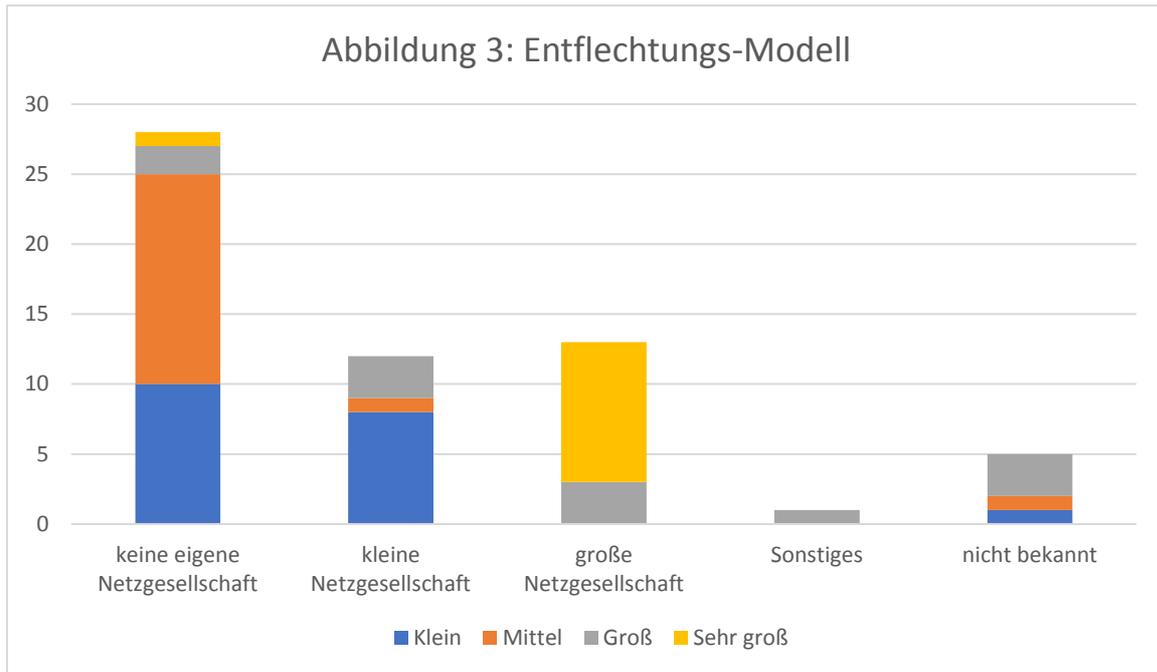

Abbildung 3: Welches Entflechtungs-Modell wurde in Ihrem Unternehmen umgesetzt?

## Teil B: Organisatorisches

In diesem Abschnitt der Umfrage wurden organisatorische Gegebenheiten abgeklärt. Darunter befanden sich Fragen um näheres zu der befragten Person zu erfahren. Zum Beispiel in welcher Abteilung er zugeordnet ist und welche Rolle er im Unternehmen innehat. Weiterhin wurden konkrete Fragen bezogen auf die IT-Sicherheit gestellt.

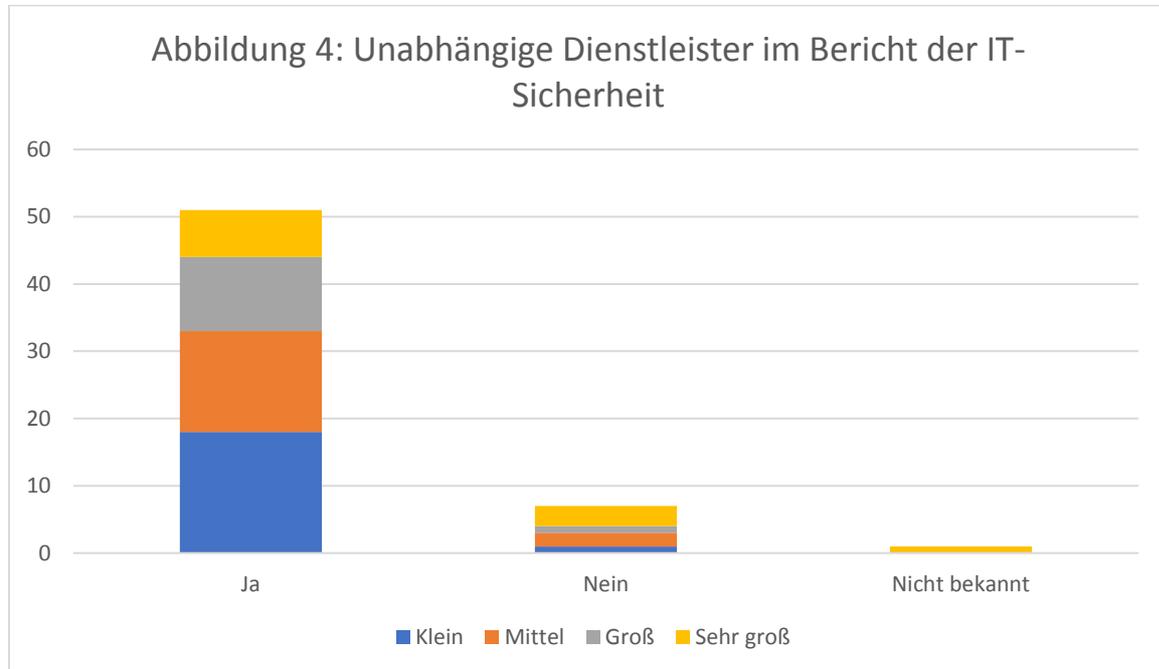

Abbildung 4: Werden in Ihrem Unternehmen, unabhängige Dienstleister im Bereich IT-Sicherheit eingesetzt?

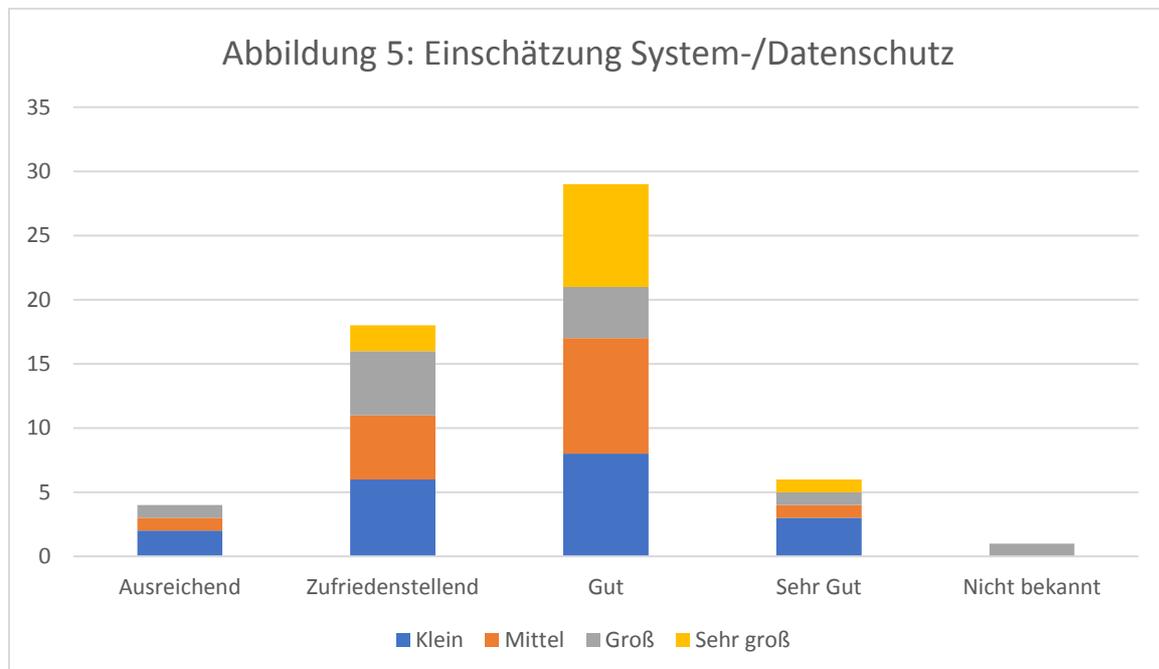

Abbildung 5: Wie gut sind Ihrer Einschätzung nach die Systeme und Daten Ihres Unternehmens geschützt?

# Teil C: Information Security Management System (ISMS)

Um einen besseren Überblick zum Status der Einführung des Information Security Management Systems zu erhalten, wurden explizit Fragen dazu gestellt.

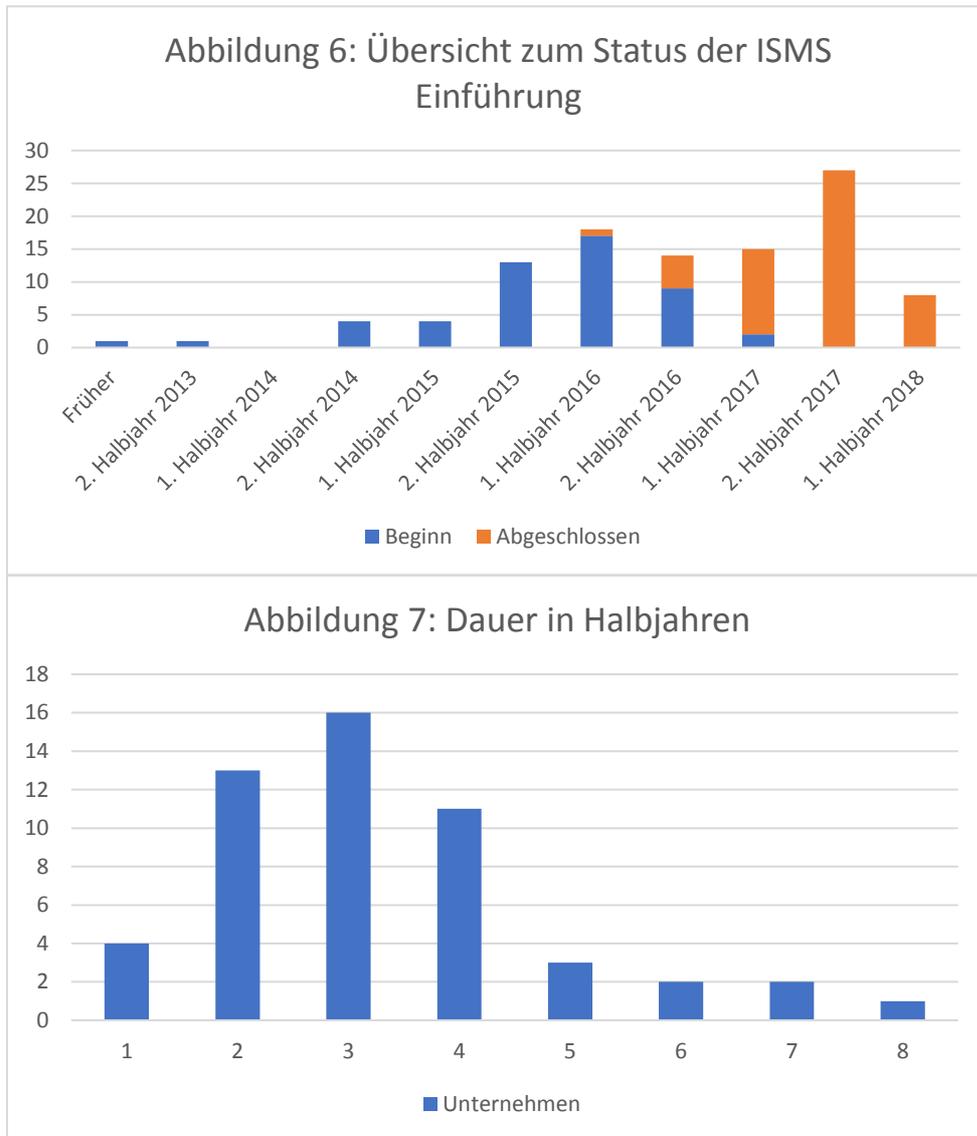

Abbildung 6 und 7 befasst sich mit folgenden Fragen:
- Die Einführung eines ISMS ...
- Wann sollen die Arbeiten zur Einführung eines ISMS beginnen?
- Wann wurde mit den Arbeiten zur Einführung eines ISMS begonnen?
- Bis wann sollen die Arbeiten zur Einführung eines ISMS abgeschlossen sein?
- Wann wurden die Arbeiten zur Einführung eines ISMS abgeschlossen?

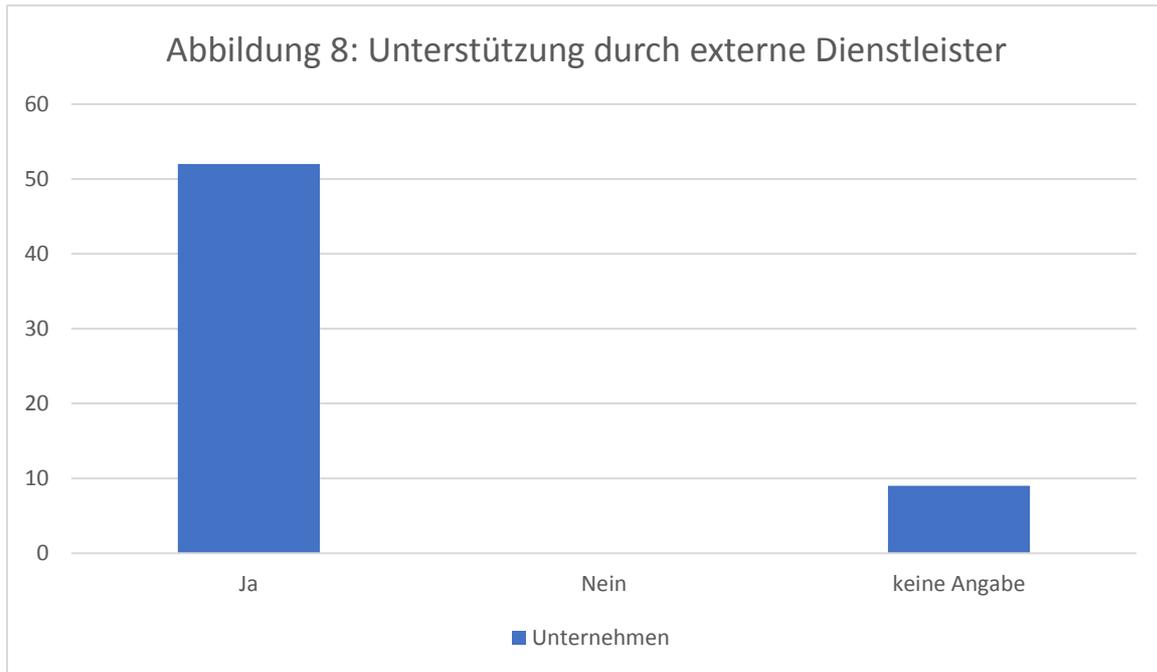

Abbildung 8: Wurden bzw. werden bei der Einführung eines ISMS externe Dienstleister (z.B. Unternehmensberater) hinzugezogen?

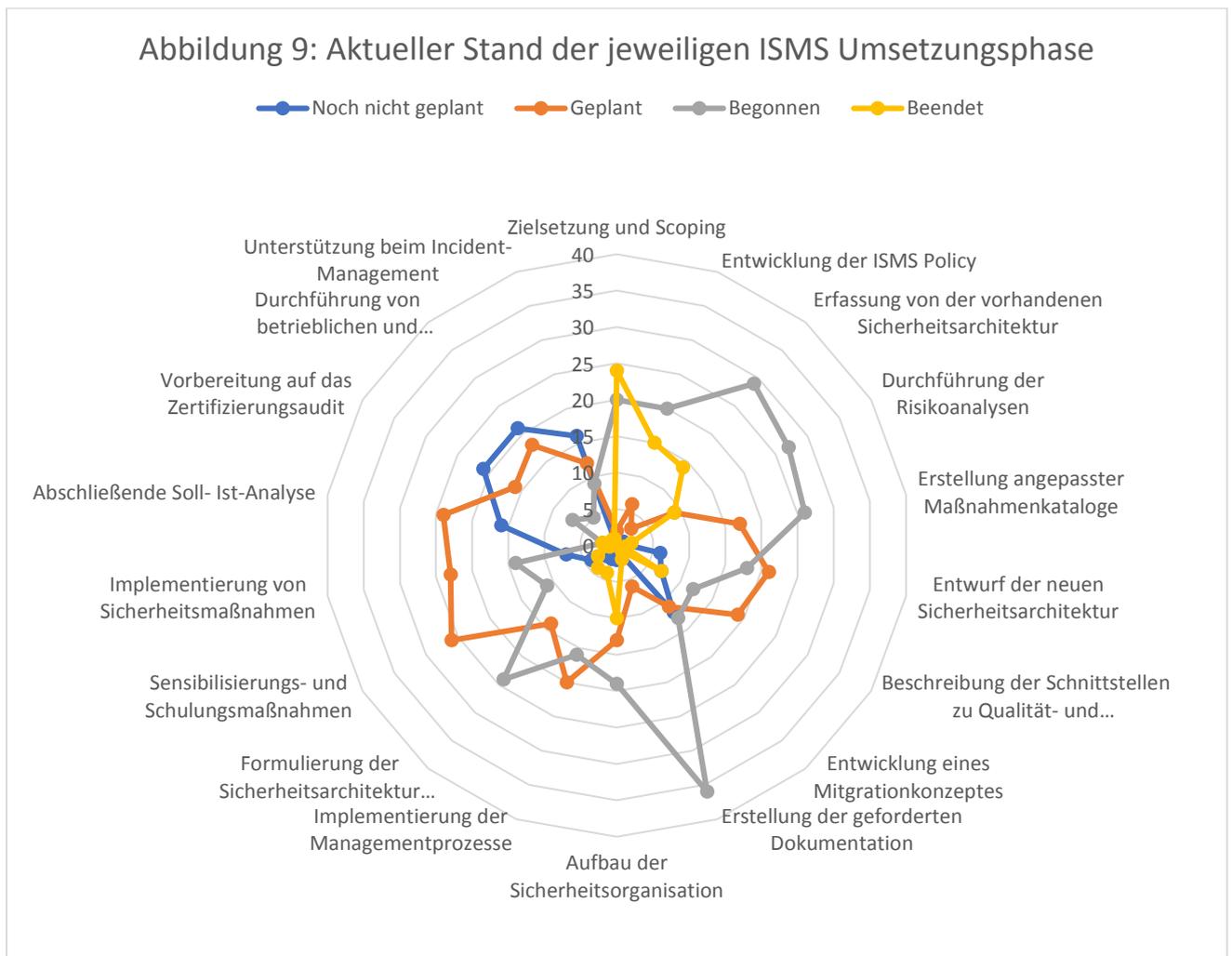

Abbildung 9: Wie ist der aktuelle Stand der jeweiligen ISMS Umsetzungsphasen?

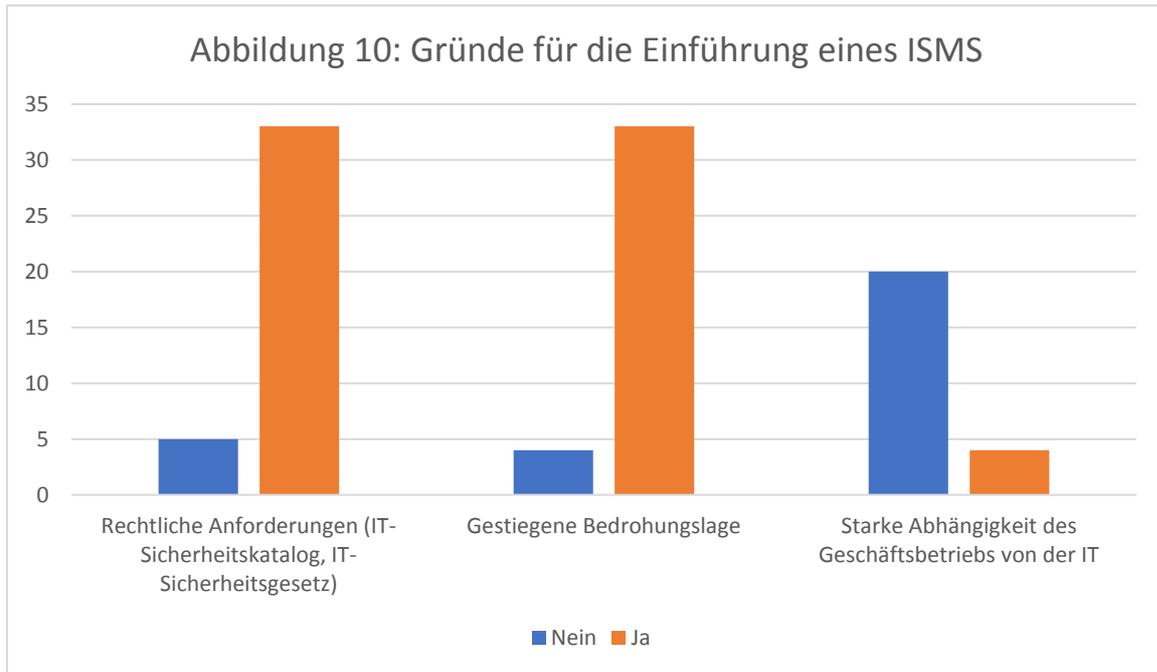

Abbildung 10: Was waren für Sie die wesentlichen Gründe für die Einführung eines ISMS (Mehrfachauswahl möglich)?

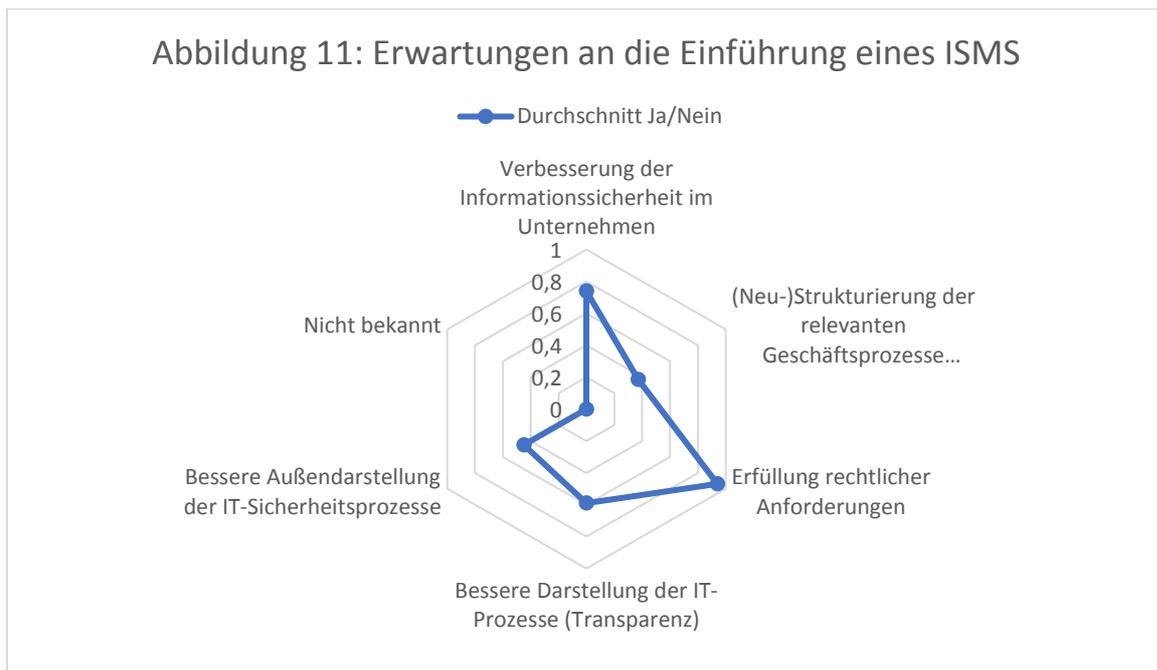

Abbildung 11: Was erhoffen Sie sich bzw. erwarten Sie von der Einführung eines ISMS (Mehrfachauswahl möglich)?

## Teil D: Büro IT

In diesem Abschnitt der Umfrage wird die Büro IT im Hinblick auf die IT-Sicherheit beleuchtet. Um eine höhere Sicherheit gewährleisten zu können muss es entsprechende IT-Sicherheitsrichtlinien geben und diese müssen regelmäßig überprüft werden. Die Überprüfung ist aufgrund der ständigen Weiterentwicklung der Technik wichtig.

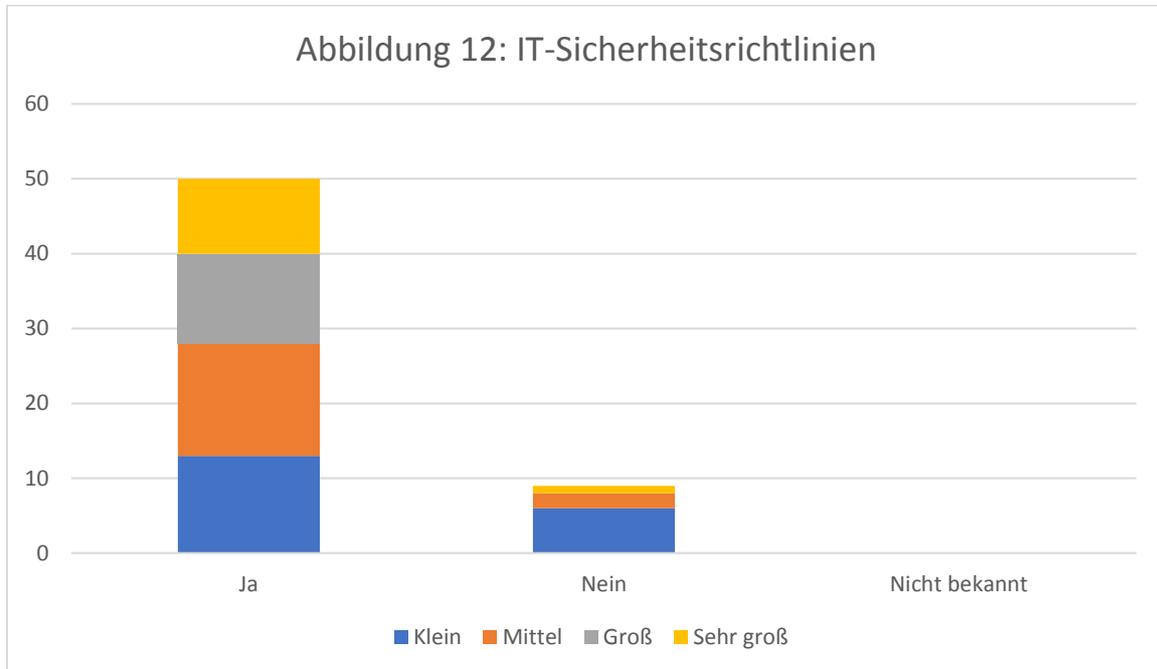

Abbildung 12: Existieren IT-Sicherheitsrichtlinien für die Büro IT Ihres Unternehmens?

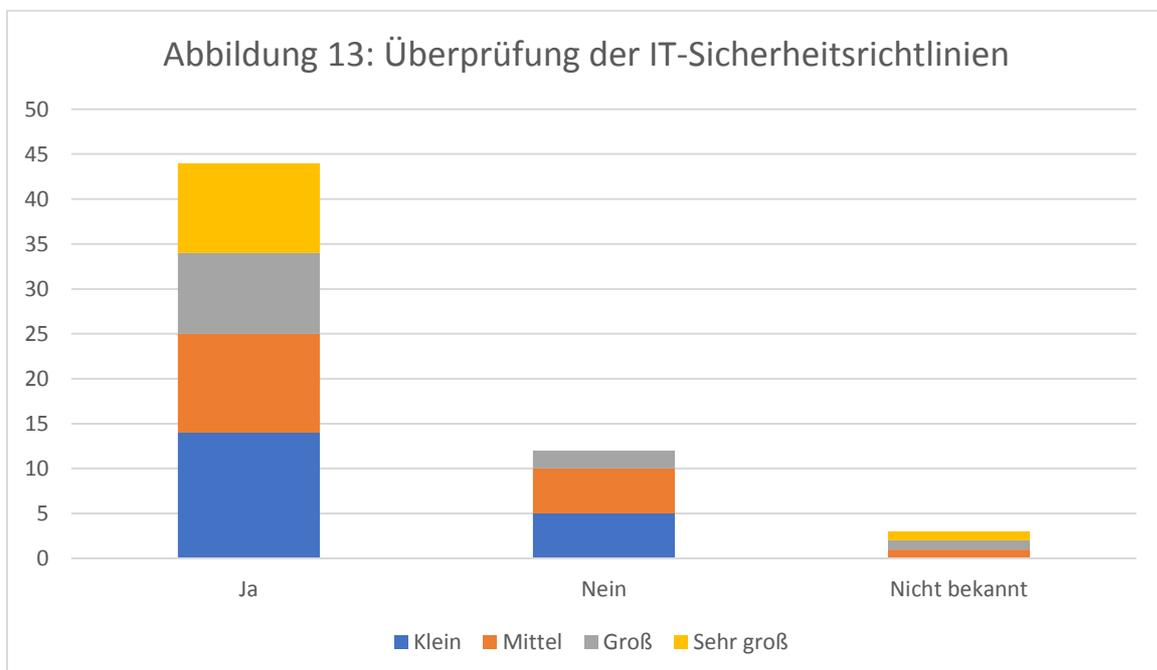

Abbildung 13: Werden die IT-Sicherheitsrichtlinien in regelmäßigen Zeitabständen überprüft und ggf. angepasst?

## Teil E: Leitsystem: Netzaufbau

Das Leitsystem stellt den Kern des Stromnetzbetreibers dar. Um mehr Informationen über den generellen Aufbau des Leitsystems zu erhalten, sind spezifische Fragen zu dem Netzaufbau gestellt worden.

Zwei Hauptaufgaben des Leitsystems sind die Netzüberwachung und –steuerung bzw. die Durchführung von Schaltvorgängen.

Ein weiterer wichtiger Aspekt des Netzaufbaus ist die Trennung zwischen dem Leitsystem und den anderen Netzwerken (z. B. Büro IT; Internet; Wartungsfirmen). Liegen keine Trennungen vor, könnte dies eine Gefahrenstelle bzw. ein möglicher Angriffspunkt sein, welcher geschützt werden muss.

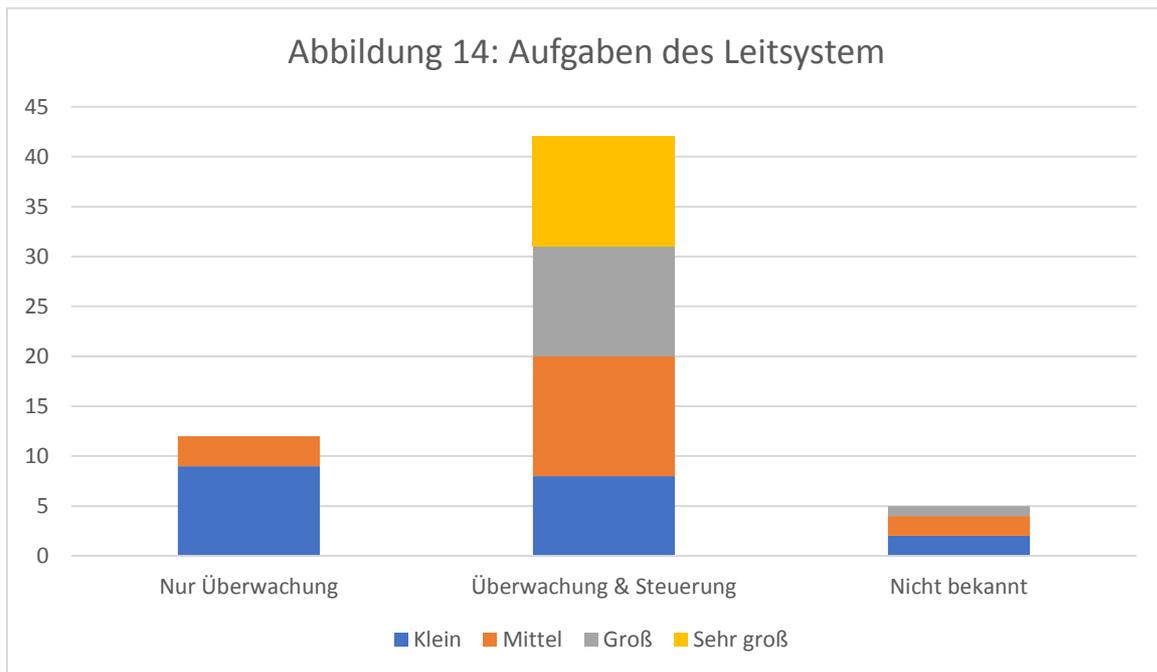

Abbildung 14: Dient Ihr Leitsystem nur der Netzüberwachung oder können hierüber auch Schaltvorgänge durchgeführt werden?

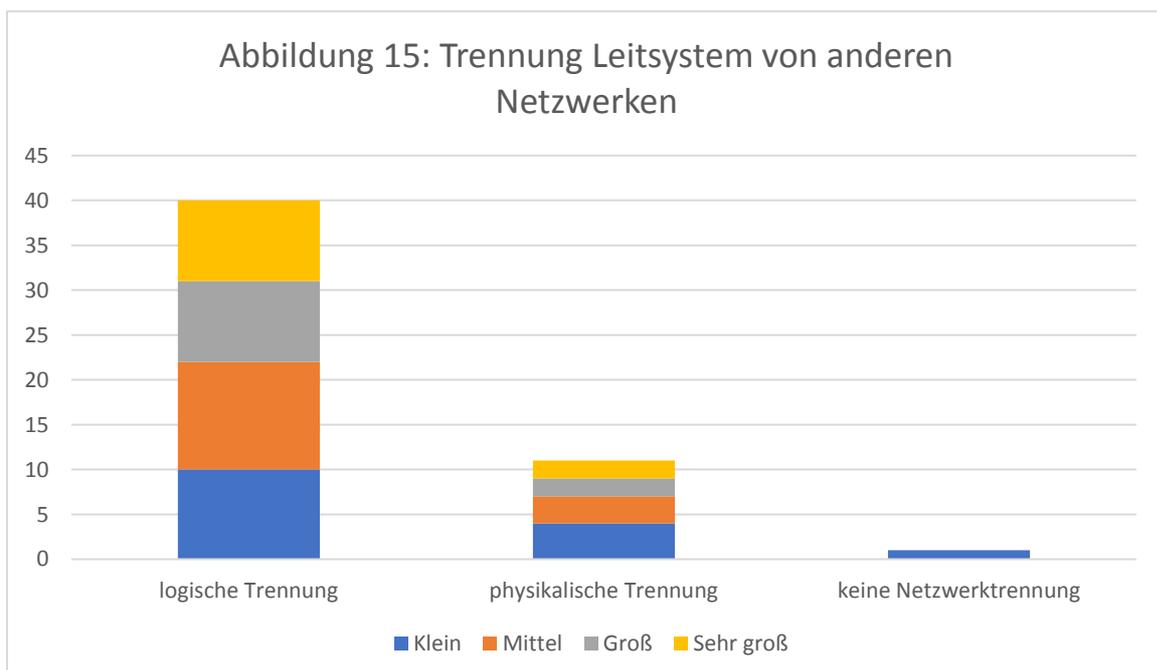

Abbildung 15: Wie ist das IT-Netzwerk Ihres Leitsystems von anderen Netzwerken (z. B. Büro IT, Internet, Wartungsfirmen) getrennt?

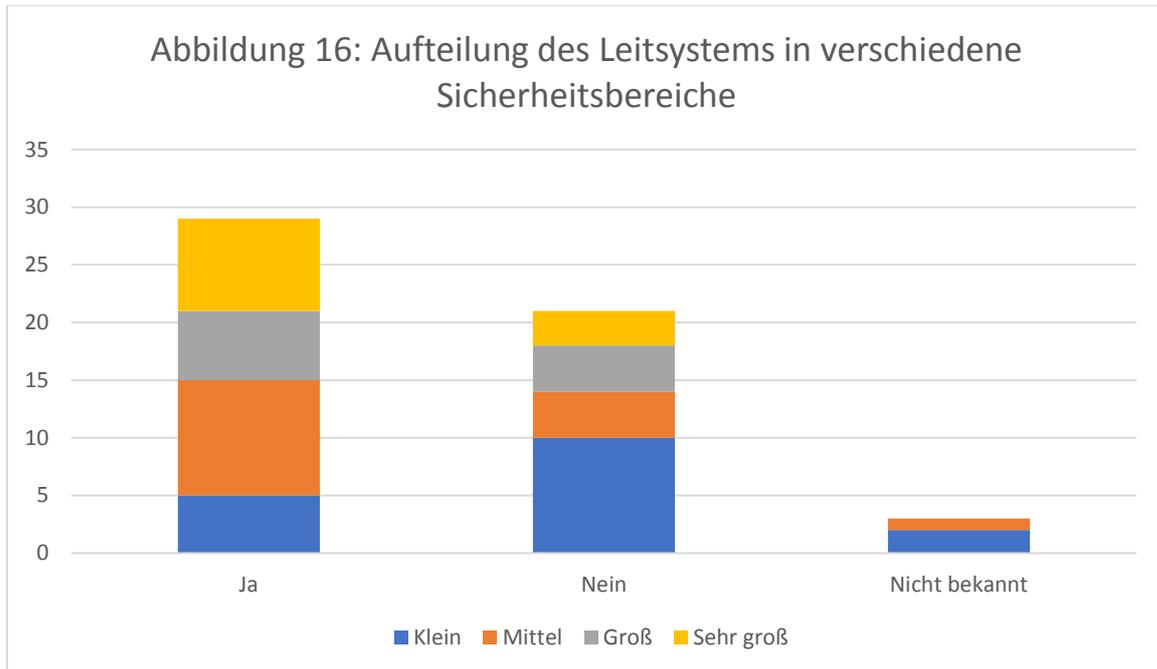

Abbildung 16: Ist das Netzwerk Ihres Leitsystems in verschiedene Sicherheitsbereiche unterteilt (z. B. durch verschiedene VLANs)?

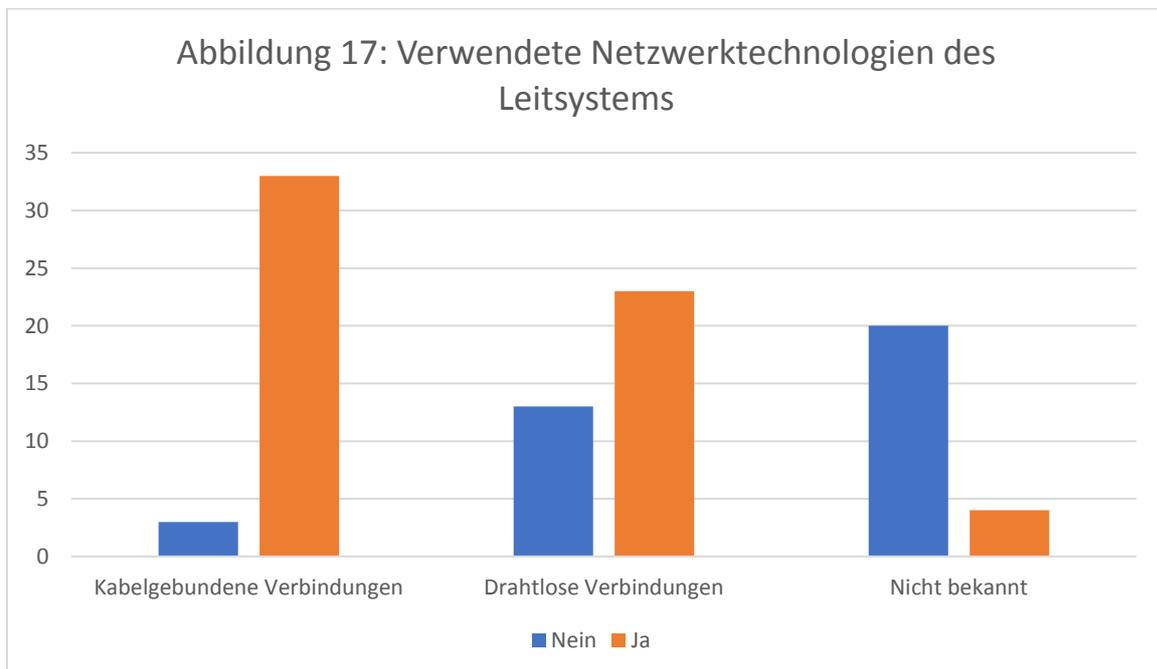

Abbildung 17: Welche Netzwerktechnologien werden im Netzwerk Ihres Leitsystems eingesetzt (Mehrfachnennungen sind möglich)?

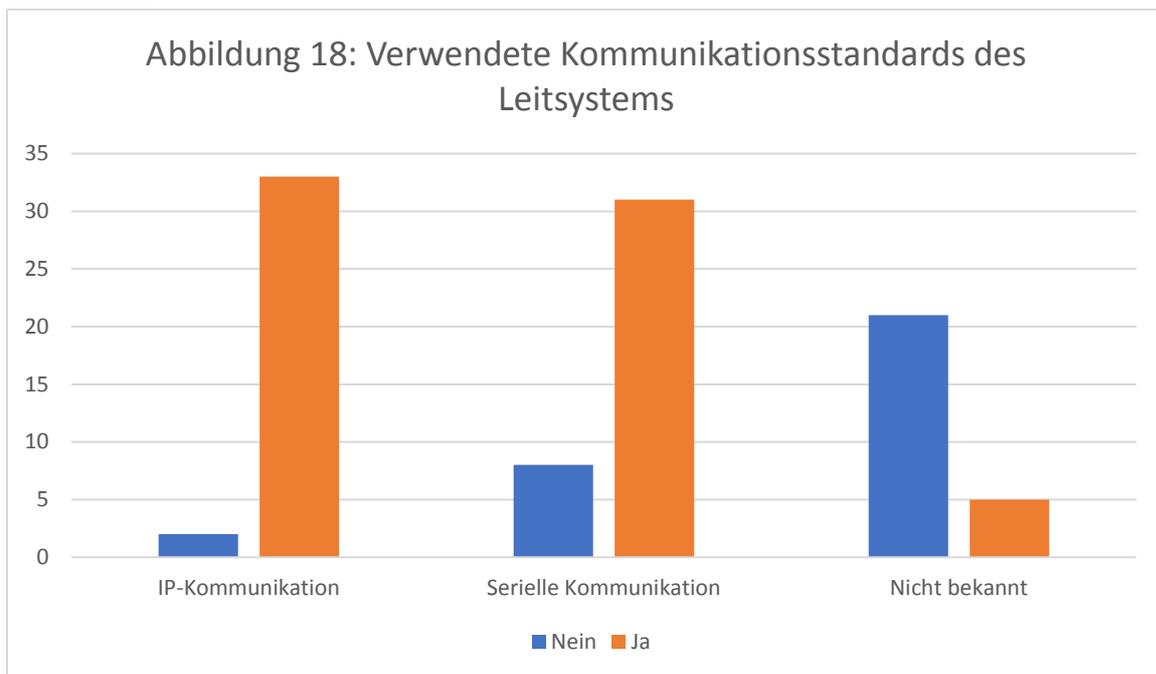

Abbildung 18: Welche Kommunikationsstandards werden im Netzwerk Ihres Leitsystems verwendet (Mehrfachnennungen sind möglich)?

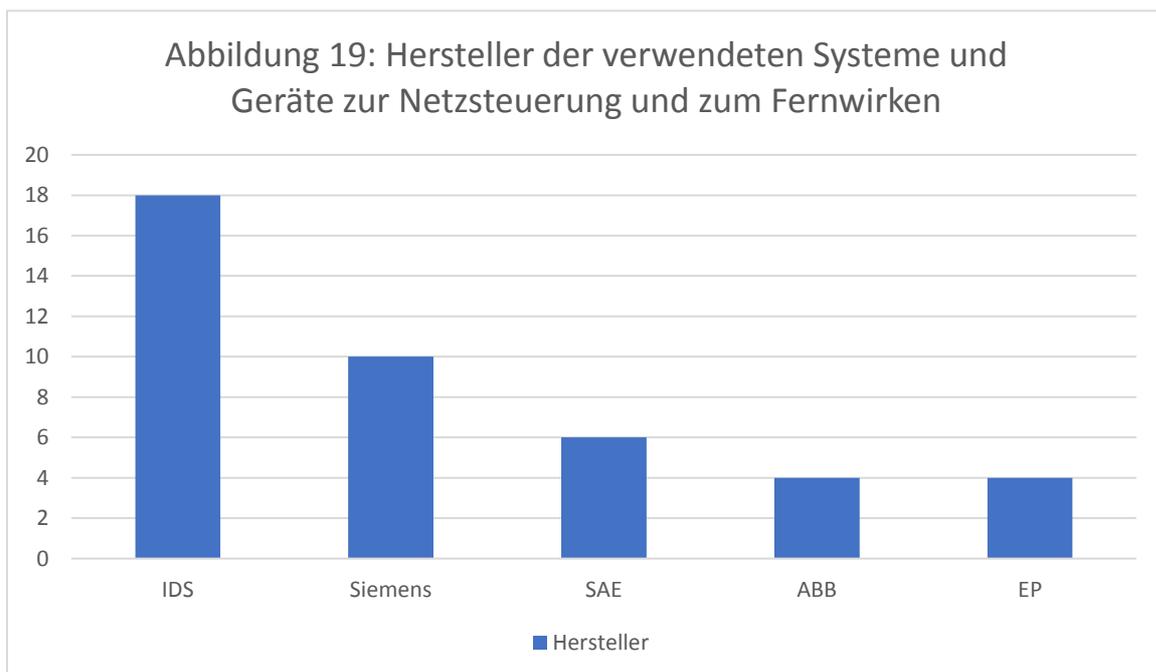

Abbildung 19: Von welchen Herstellern setzen Sie Systeme und Geräte zur Netzsteuerung und zum Fernwirken ein? (ab mind. vier Nennungen in der Abbildung aufgenommen)

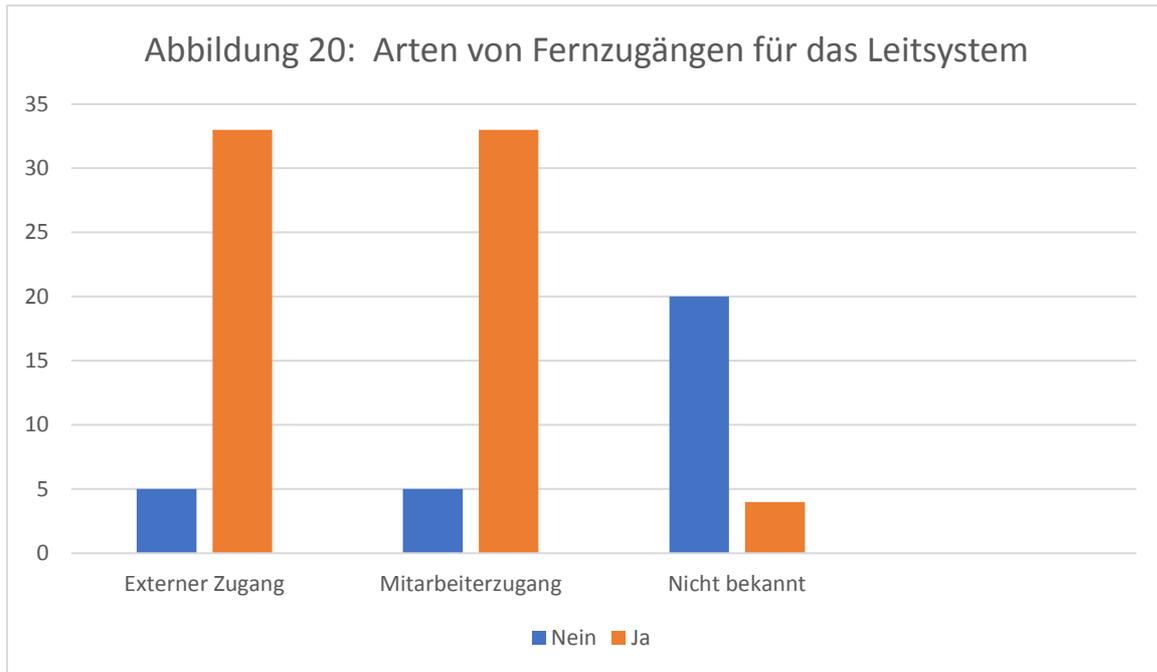

Abbildung 20: Welche Arten von Fernzugängen sind für Ihr Leitsystem eingerichtet (Mehrfachnennungen möglich)?

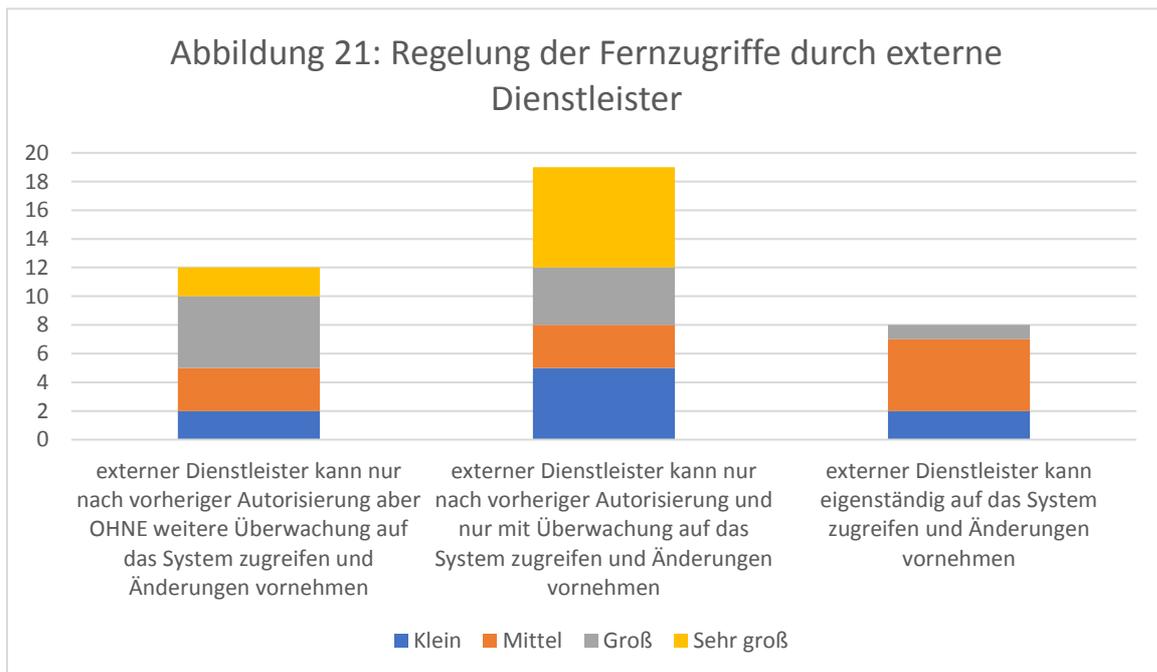

Abbildung 21: Wie sind Fernzugriffe durch externe Dienstleister geregelt?

## Teil F: Leitsystem: Prozess und Organisation

Neben den technischen Daten des Leitsystems, ist das Prozess und die organisatorischen Strukturen dahinter mindestens genauso wichtig. IT-Sicherheit muss stetig überwacht und verbessert werden, da sich die Mittel und Techniken der potentiellen Angreifer ständig weiterentwickelt. Genauso müssen aufgedeckte Schwachstellen behandelt werden und es sollte eine regelmäßige Überwachung und Informationsweitergabe innerhalb des Unternehmens gegeben sein, um eine Prävention gegen Hackerangriffe bieten zu können.

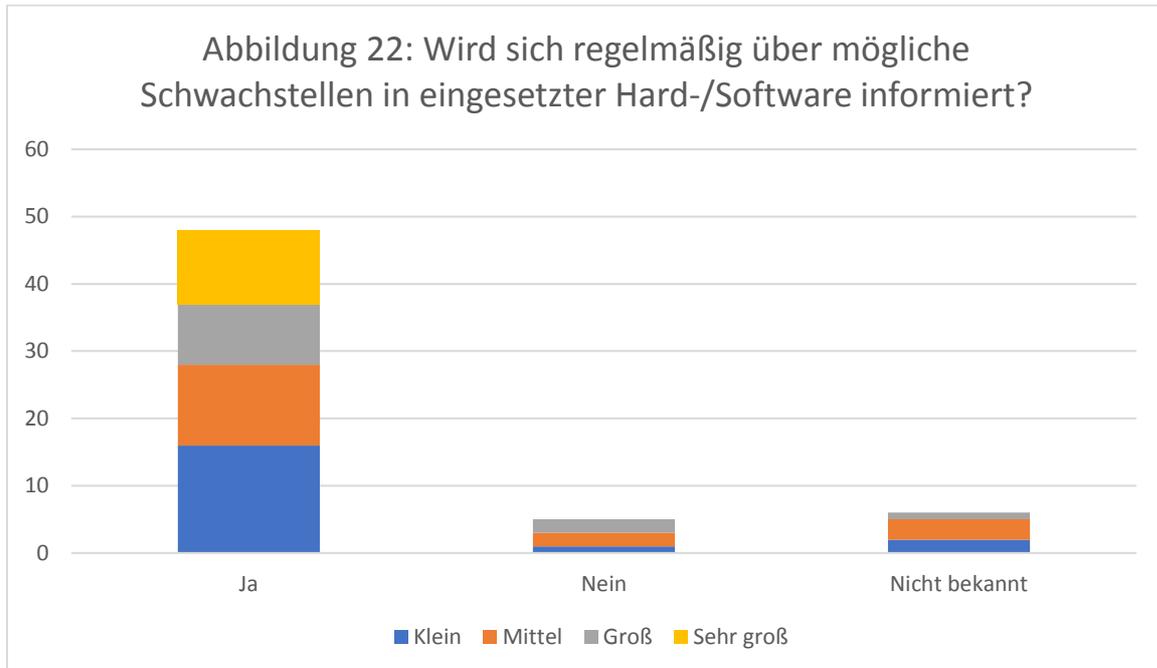

Abbildung 22: Informieren Sie sich, bzw. die verantwortlichen Mitarbeiter Ihres Unternehmens, regelmäßig über mögliche Schwachstellen eingesetzter Hard- und Software?

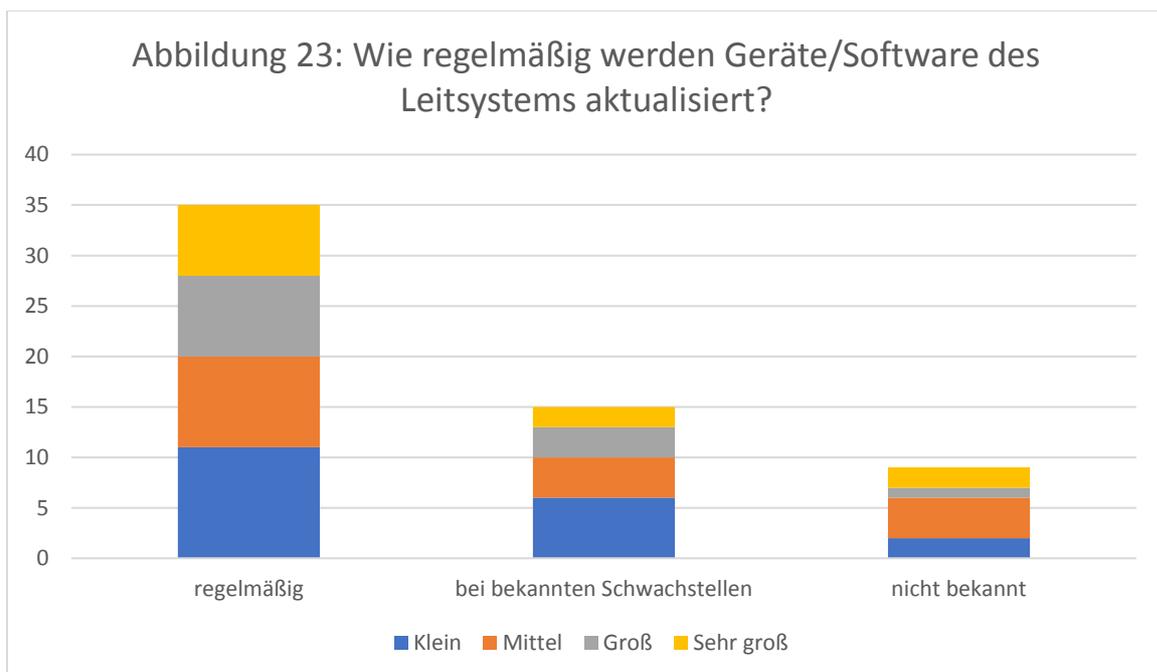

Abbildung 23: Wie regelmäßig werden Geräte und Software innerhalb Ihres Leitsystems aktualisiert bzw. erneuert?

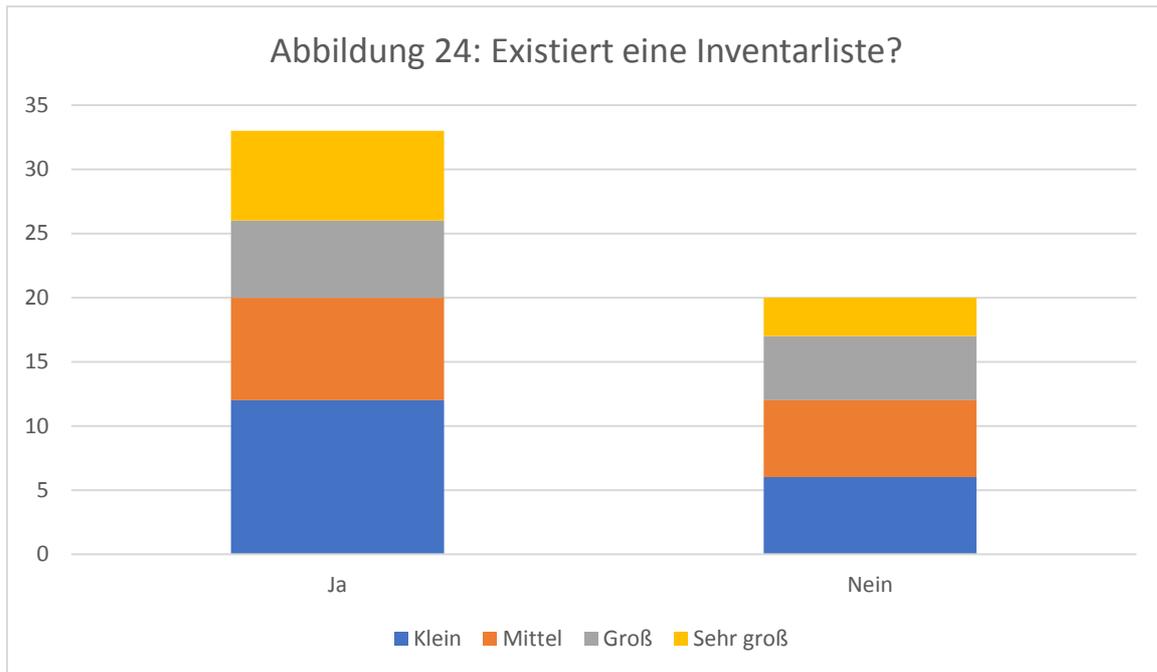

Abbildung 24: Existiert eine aktuelle Inventarliste, in der alle Softwarestände dokumentiert sind (z. B. mit Versionsnummern, zugeordneten Accounts und IP-Adressen)?

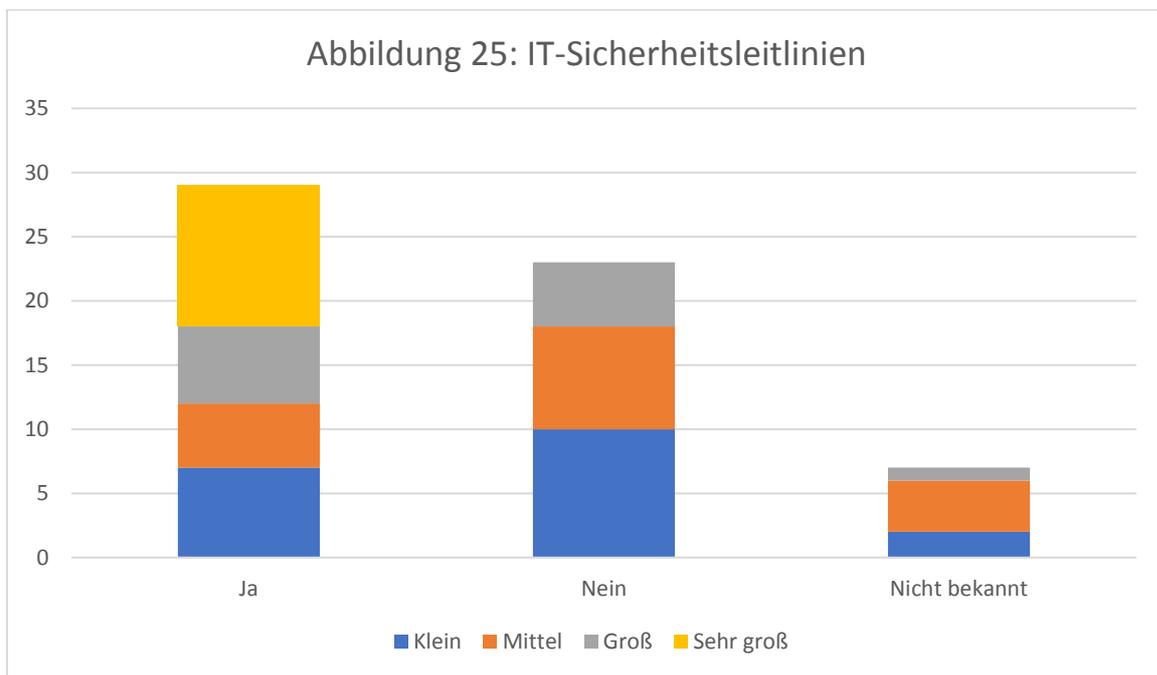

Abbildung 25: Gibt es in Ihrem Unternehmen niedergeschriebene IT-Sicherheitsleitlinien für den Bereich des Leitsystems?

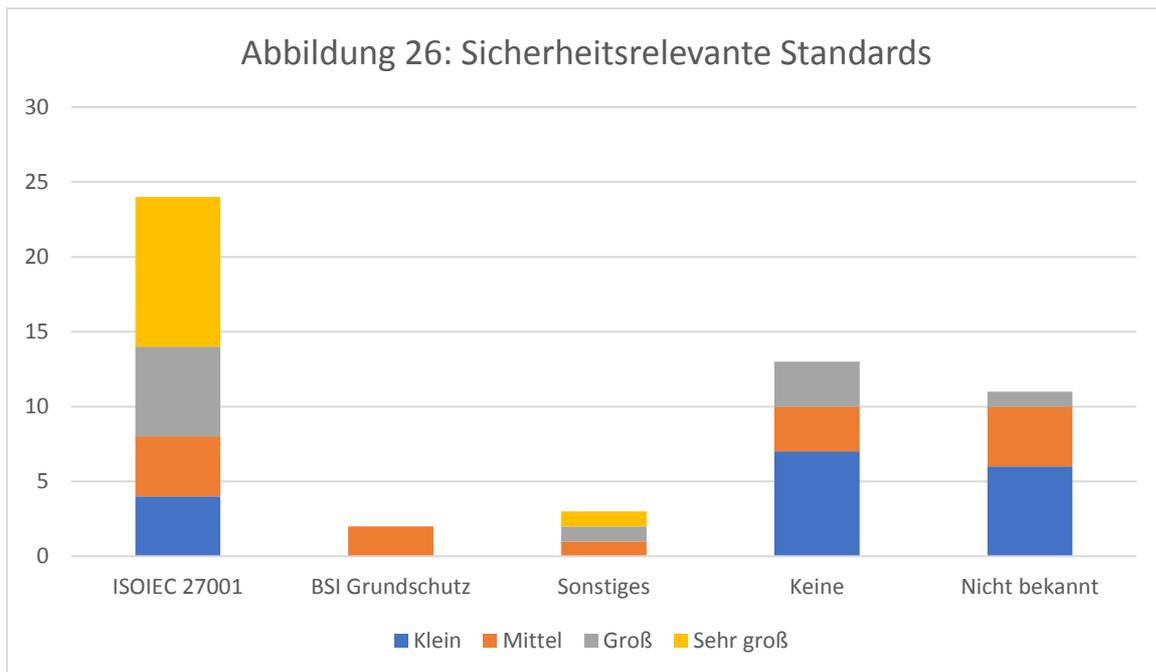

Abbildung 26: Anhand welcher sicherheitsrelevanter Standards sind Ihre IT-Systeme und Prozesse zu Netzsteuerung ausgelegt?

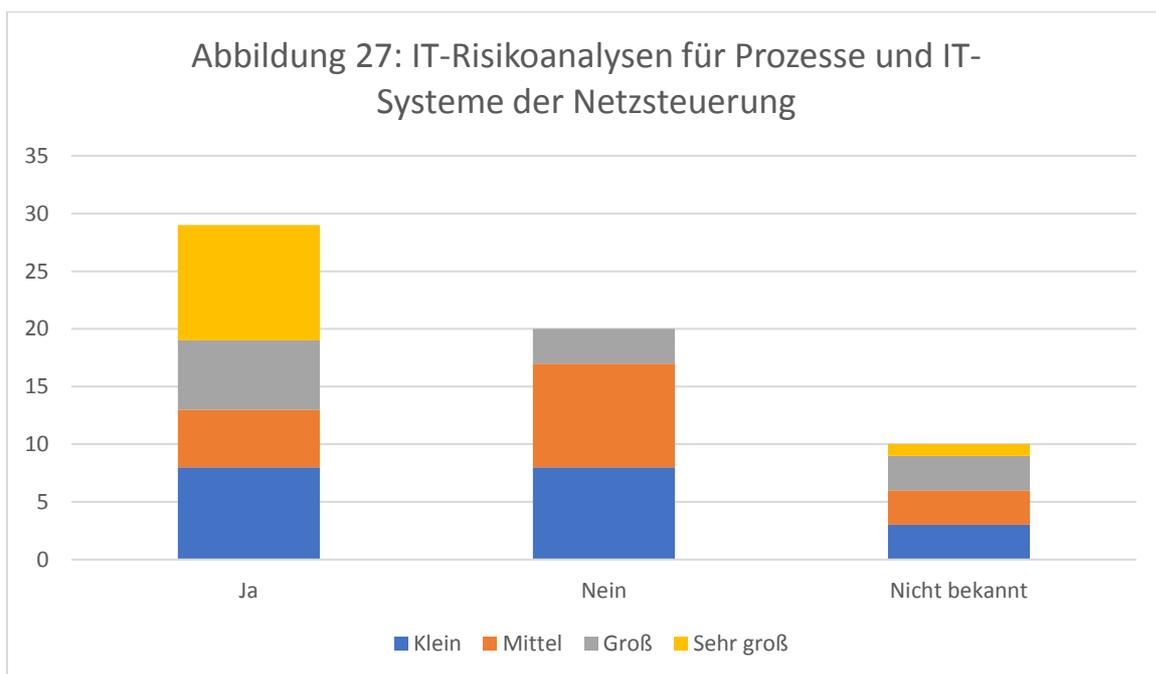

Abbildung 27: Führen Sie IT-Risikoanalysen für die Prozesse und IT-Systeme zur Netzsteuerung durch?

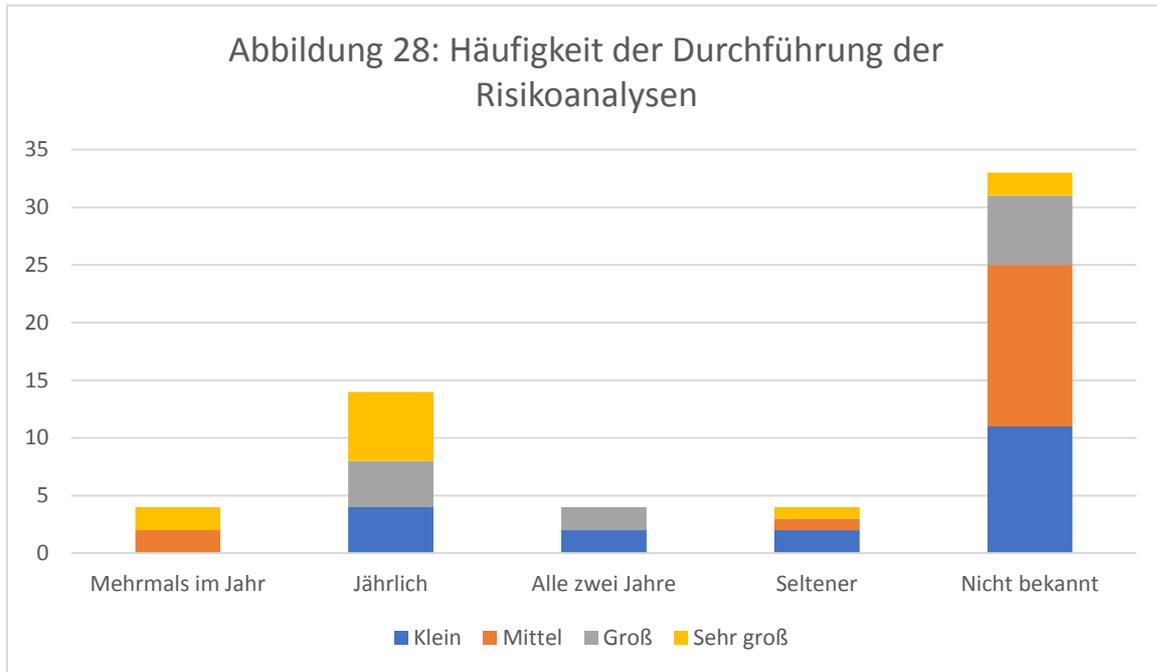

Abbildung 28: Wie regelmäßig führen Sie solche Risikoanalysen durch?

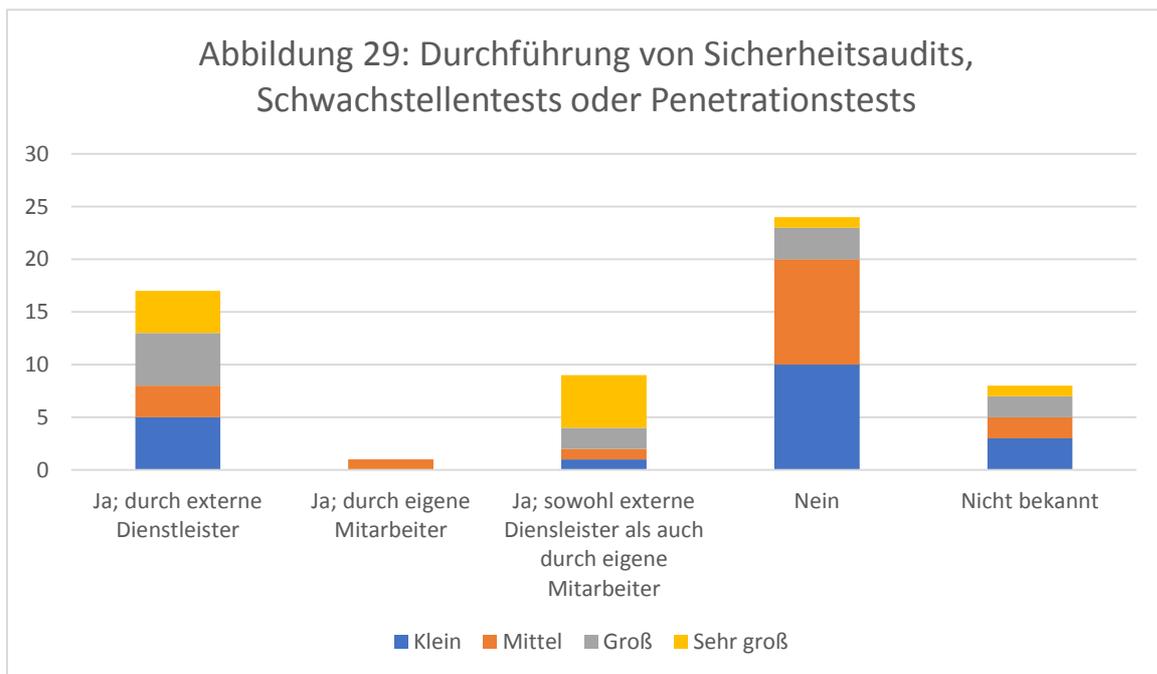

Abbildung 29: Führen Sie Sicherheitsaudits, Schwachstellenscans oder Penetrationstests für die Systeme zur Steuerung der Netzleittechnik durch?

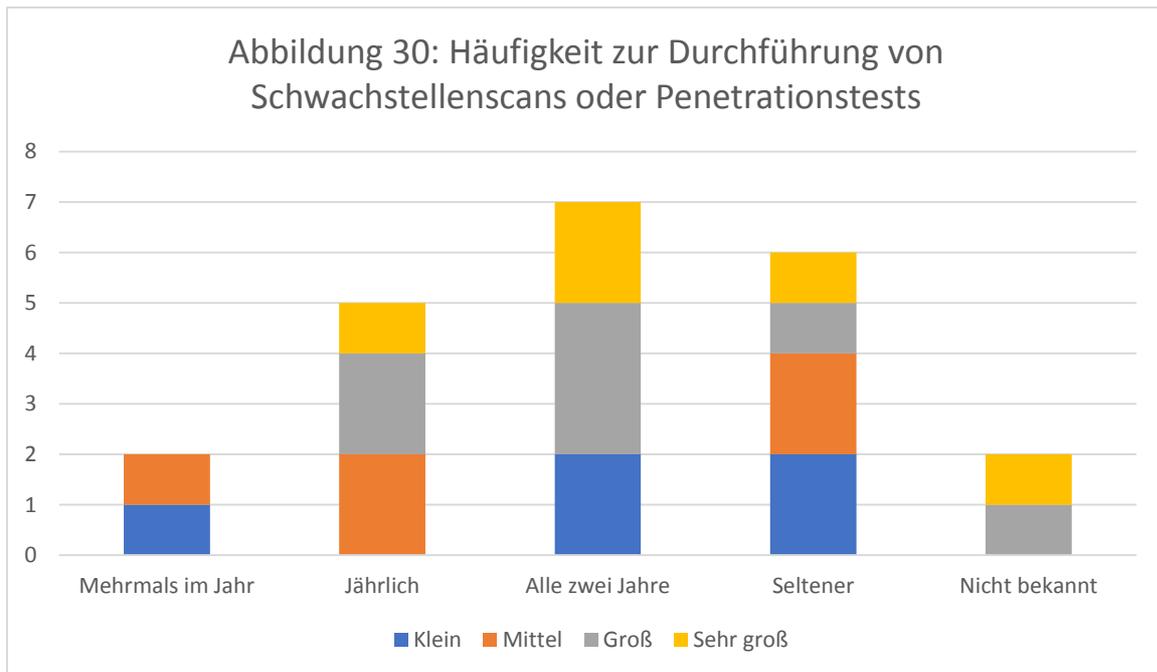

Abbildung 30: Wie häufig führen Sie solche Schwachstellenscans oder Penetrationstests durch?

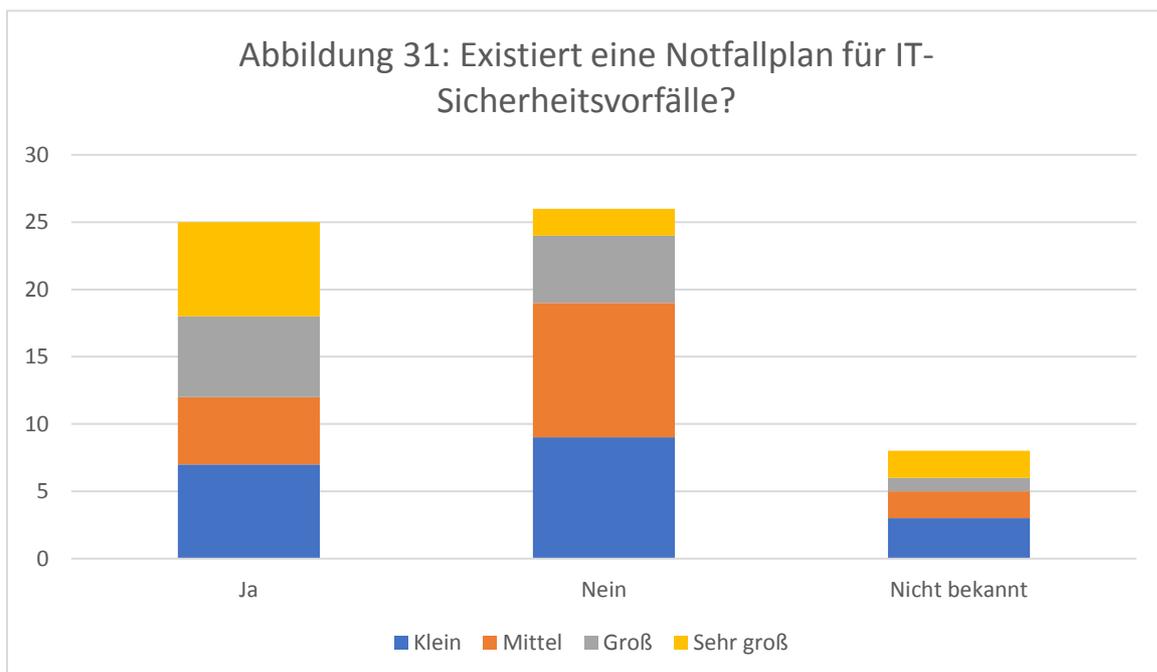

Abbildung 31: Haben Sie einen Notfallplan für IT-Sicherheitsvorfälle die die Netzsteuerung betreffen?

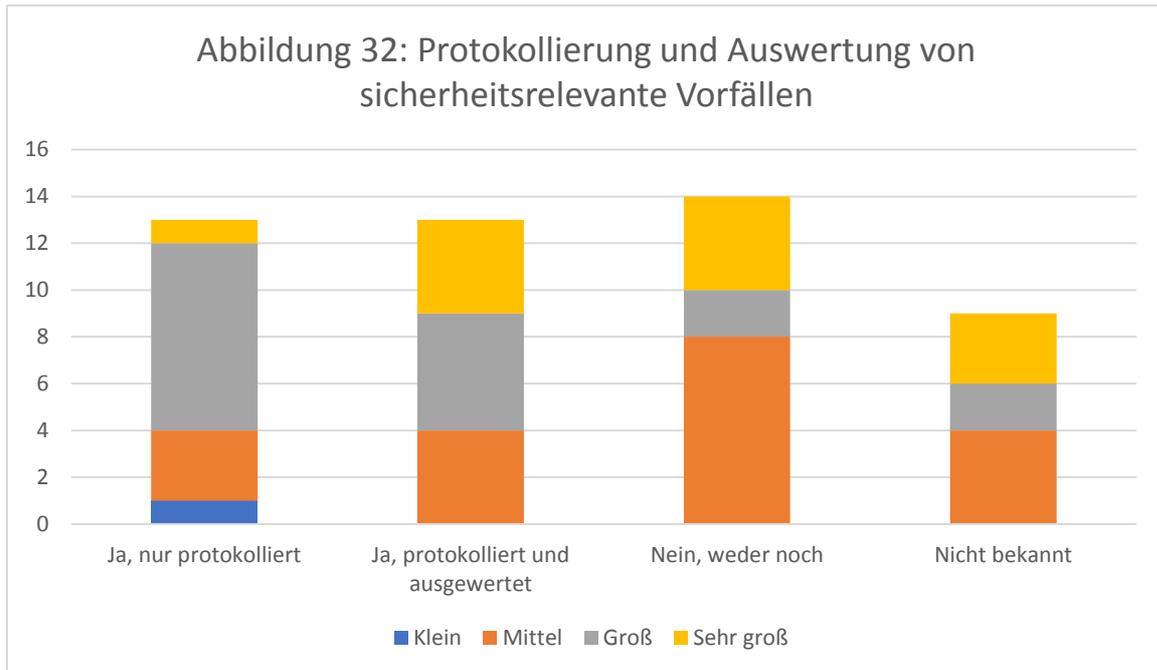

Abbildung 32: Werden sicherheitsrelevante Vorfälle (z. B. Portscans, fehlgeschlagene Anmeldeversuche, nicht autorisierte Vorgänge) protokolliert und ausgewertet?

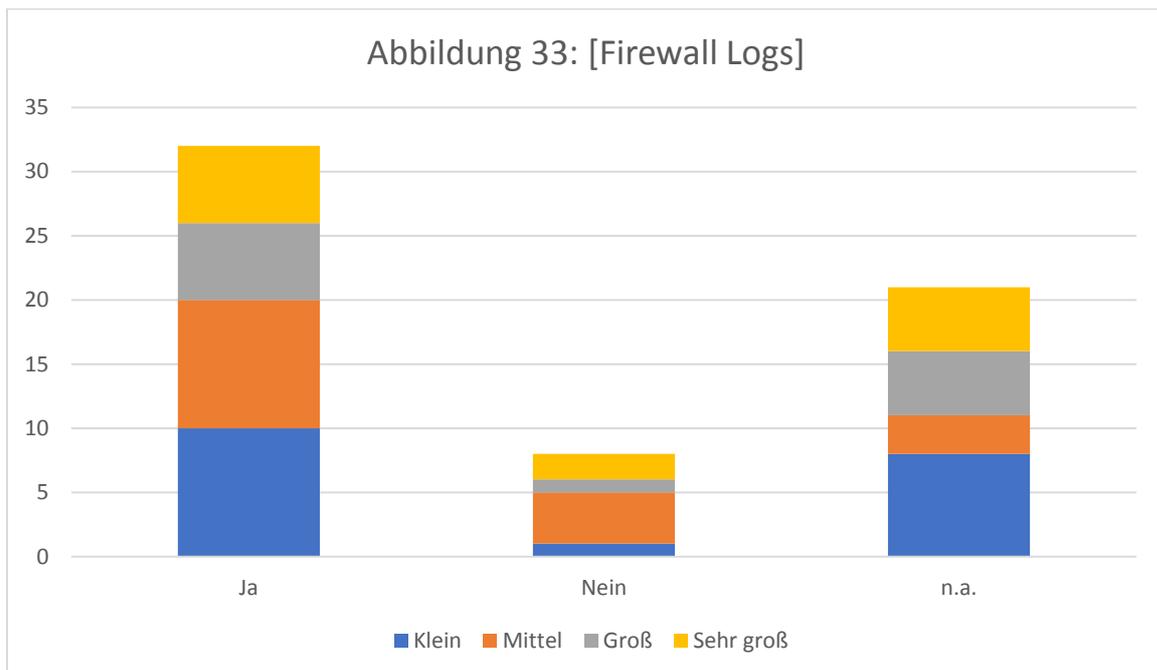

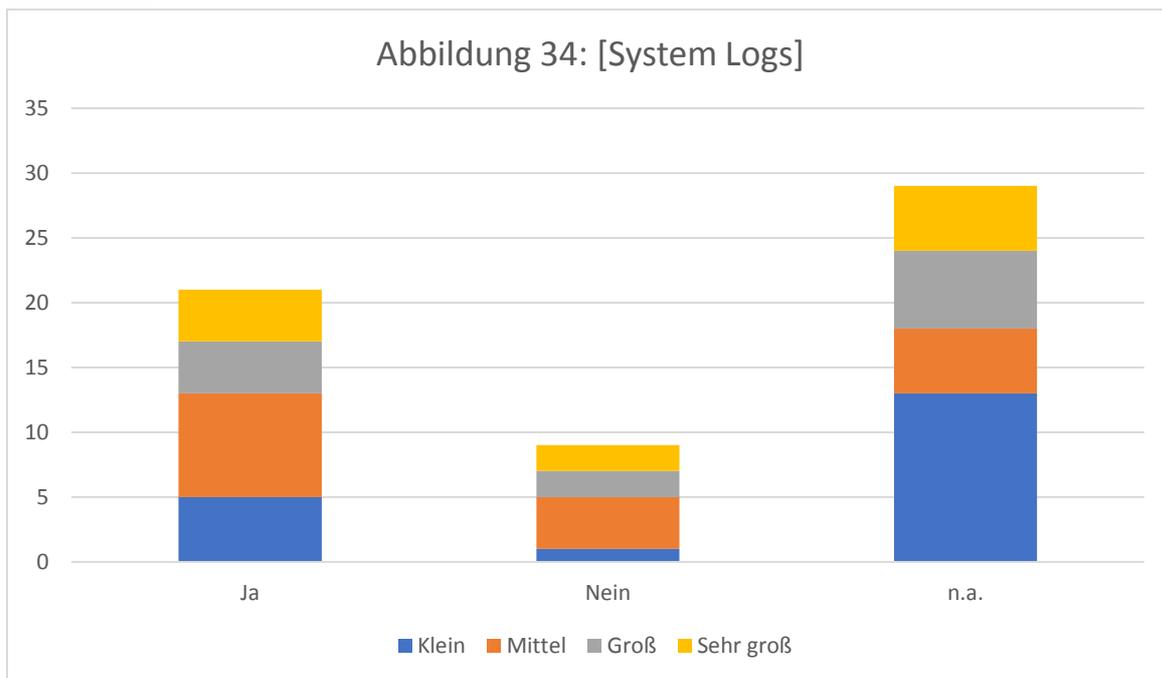

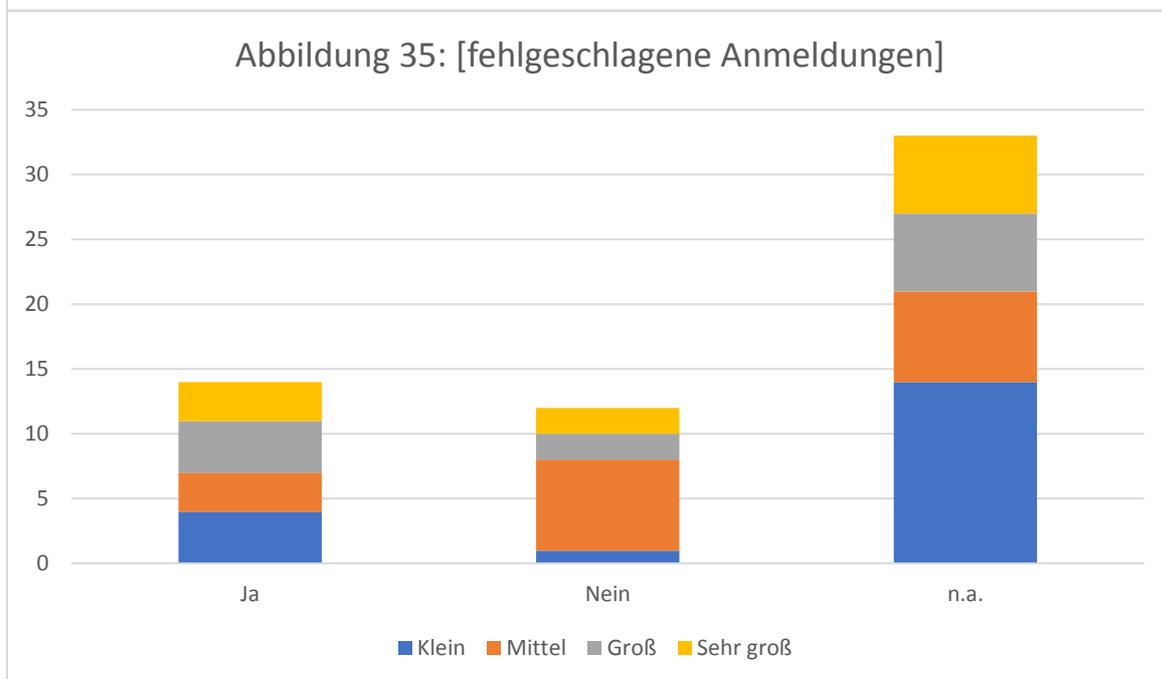

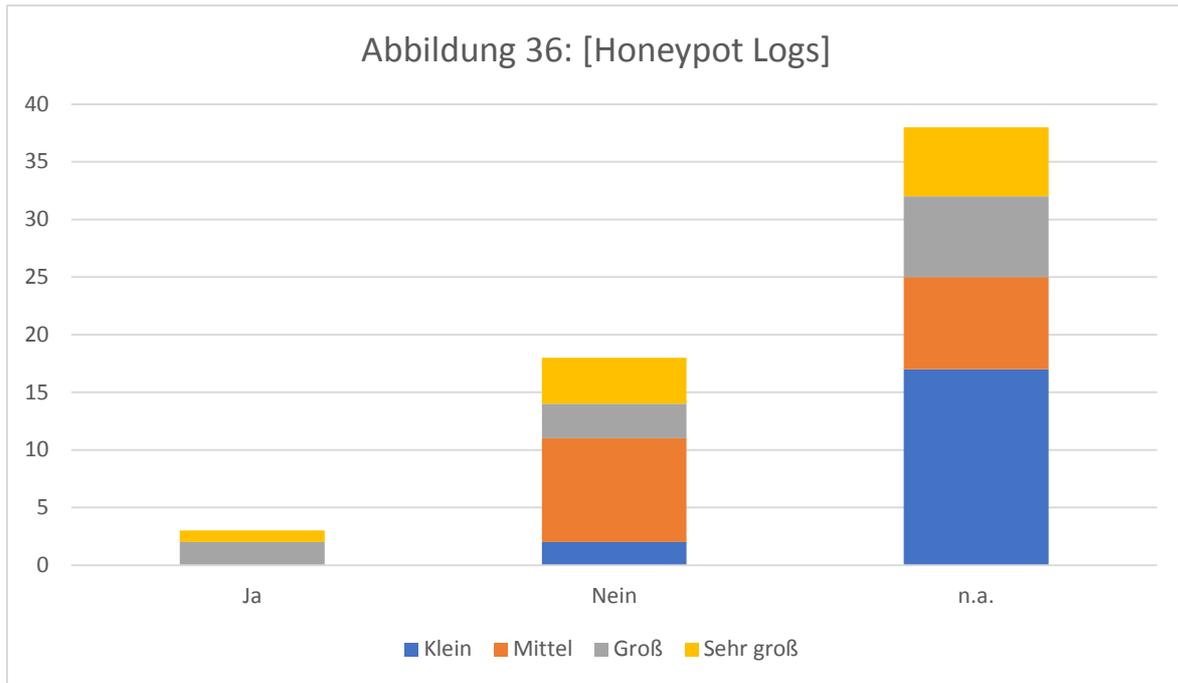

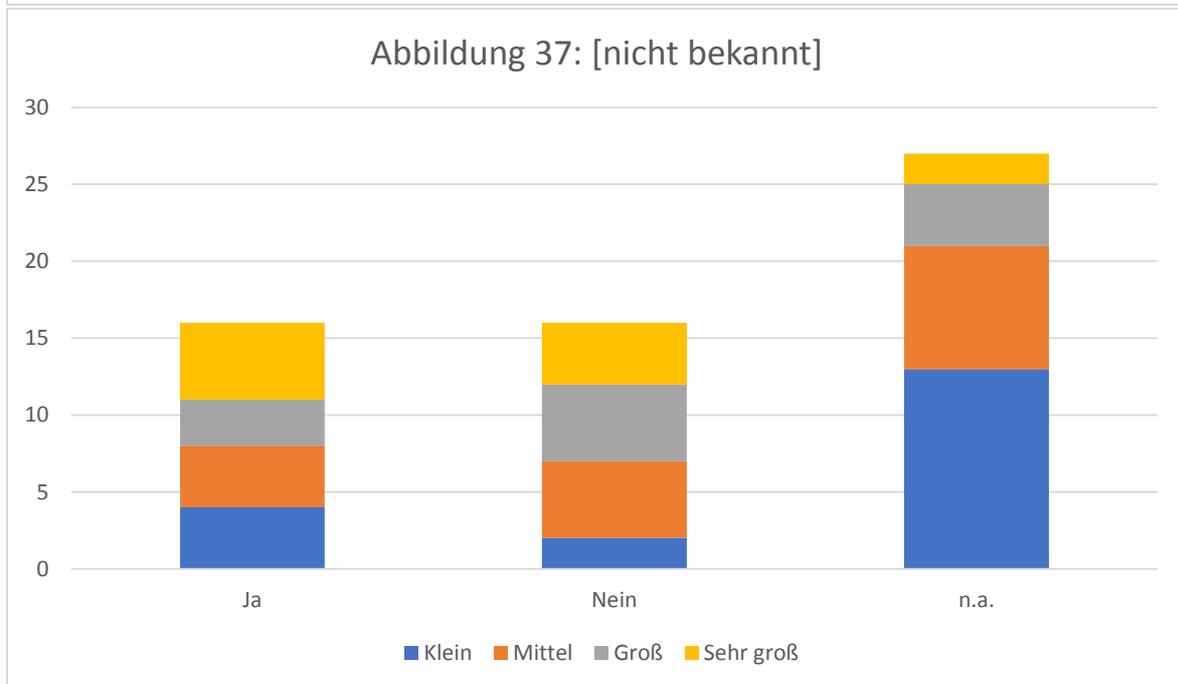

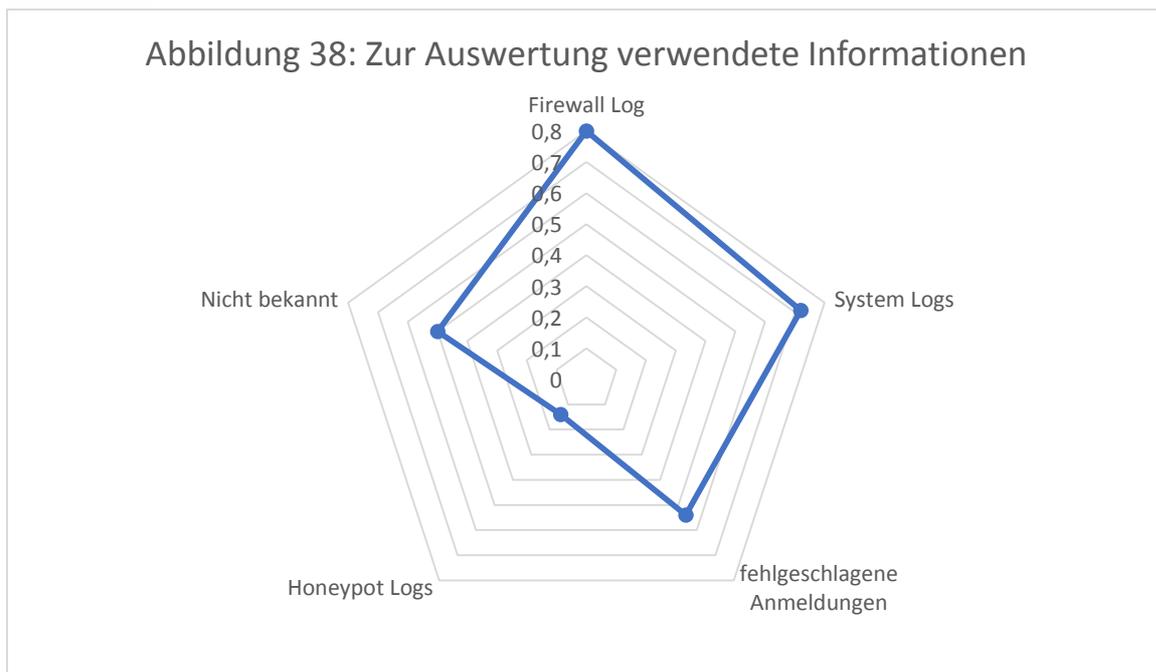

Abbildung 33 bis 38: Welche Informationen werten Sie zur Identifikation von Angriffen auf die IT-Systeme zur Netzsteuerung aus (Mehrfachauswahl möglich)?

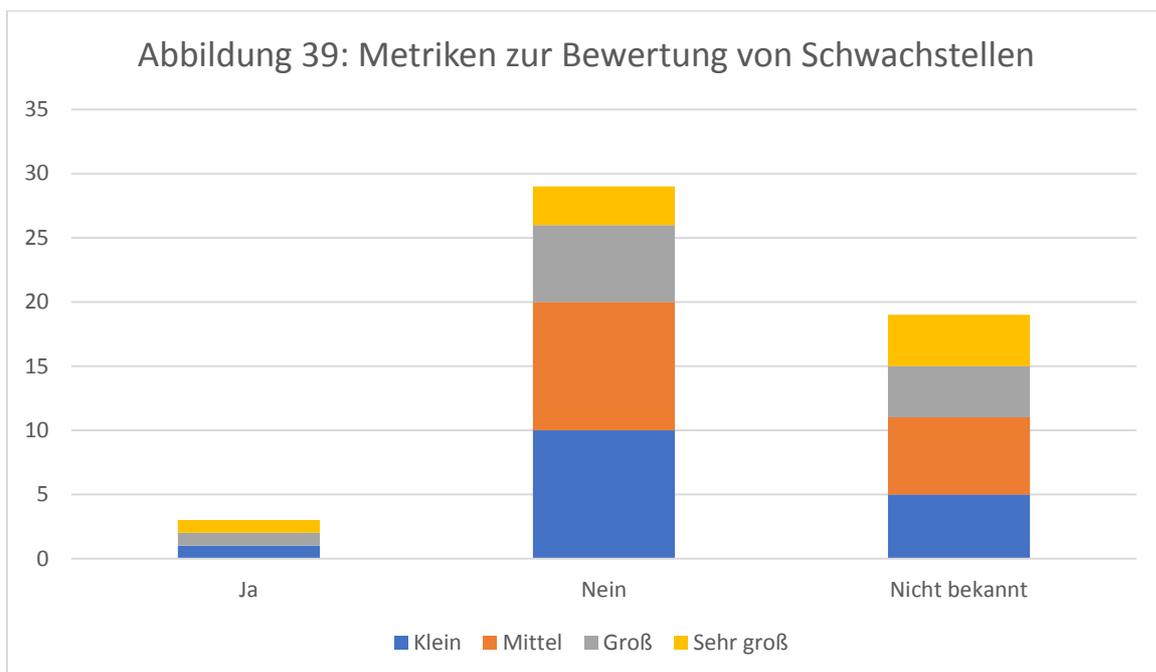

Abbildung 39: Setzen Sie Metriken zur Bewertung von Schwachstellen ein (z. B. CVSS)?

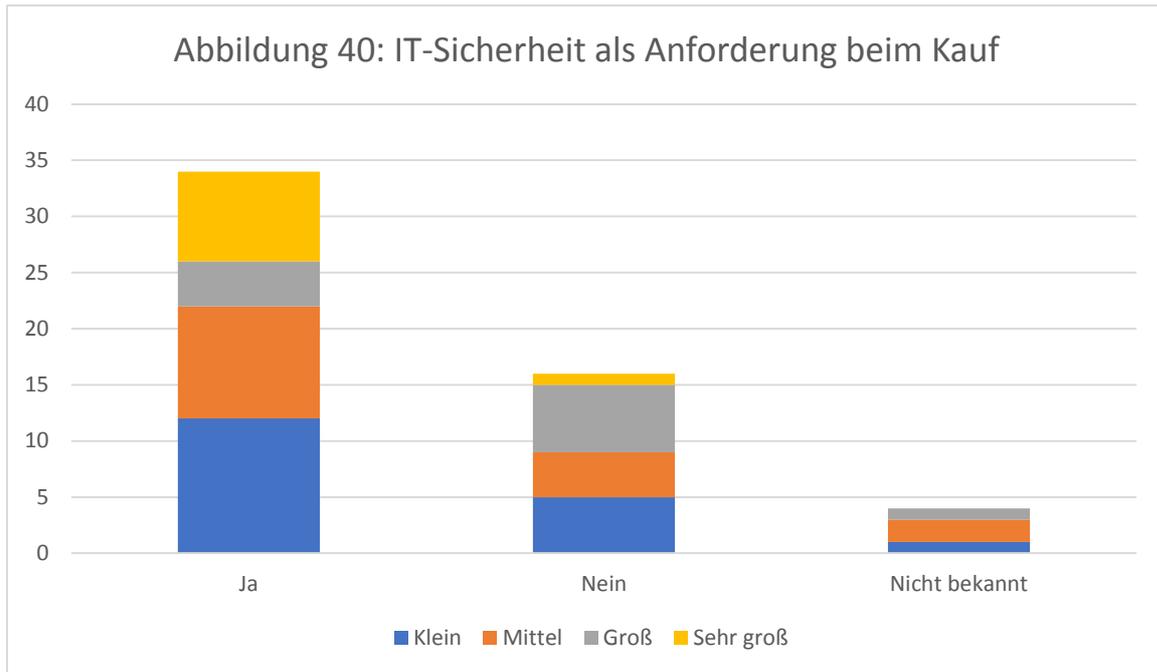

Abbildung 40: Ist IT-Sicherheit als eine Anforderung beim Kauf neuer Hard- und Software definiert?